%% file: published.tex
\documentstyle[aps,pra,multicol,amssymb,epsf,float]{revtex}

\newcommand{\id}{{\sf 1 \hspace{-0.3ex} \rule{0.1ex}{1.52ex}
\rule[-.01ex]{0.3ex}{0.1ex}}}
\newcommand{\proy}{I\!\!P}

\title{Cavity Loss Induced Generation of Entangled Atoms}
\author{M.B. Plenio, S.F. Huelga, A. Beige and P.L. Knight}
\address{Optics Section, Blackett Laboratory, Imperial College London,
London SW7 2BZ, England}
\date{\today}

\begin{document}
\draft
\maketitle
\begin{abstract}
We discuss the generation of entangled states of two two-level atoms inside an
optical resonator. When the cavity decay is continuously monitored, the absence
of photon-counts is associated with the presence of an atomic entangled state.
In addition to being conceptually simple, this scheme could be demonstrated with presently available technology. We describe how such a state is generated through
conditional dynamics, using quantum jump methods, including both cavity damping
and spontaneous emission decay, and evaluate the fidelity and relative entropy
of entanglement of the generated state compared with the target entangled state.
\end{abstract}

\vspace*{0.2cm} \noindent 

\begin{multicols}{2}

\section{Introduction}
Superposition effects in composite systems are well known in classical physics.
However, when the superposition principle is combined with a tensor product
structure for the space of states, an entirely quantum mechanical effect
arises: Quantum states can be entangled \cite{PLE98a}. This fact was early
recognised as {\em the} characteristic of the quantum formalism \cite{epr}. However,
early work concentrated on the implications of entanglement on the non local
structure of quantum theory \cite{bell} and it was considered by many as a
purely philosophical issue. The reason for the renewed interest in the fundamental
aspects of Quantum Mechanics is twofold. On one hand, it was discovered that
Bell's inequalities do not provide a good criterion for discriminating between
classical and quantum correlations when dealing with mixed states \cite{qi1}.
New criteria for characterising the separability of a given quantum state have
been proposed \cite{qi2} and measures of entanglement have been introduced \cite{qi3,matpra}.
On the other hand, it has been realized that entangled
states allow new practical applications, ranging from quantum computation
\cite{qi4} and secure cryptographic schemes \cite{qi4a} to improved
optical frequency standards \cite{susana}. The feasibility of some these applications
has been demonstrated in recent experiments\cite{qi5}. In particular, recent
advances in ion trapping technology \cite{rewWineland}
and cavity QED \cite{haroche} provide suitable scenarios for manipulating small
quantum systems.

In this paper we will discuss a scheme that allows the generation of
a maximally entangled state of two two-level atoms within a single mode cavity field.
The underlying idea is conceptually simple and relies on the concept of conditional
dynamics due to continuous observation of the cavity field. The key to understanding
how the entangled state is generated in this scheme is population trapping
\cite{knight}. There are three dressed states of the combined two-atom plus cavity
field mode system; one has a zero eigenvalue, which is therefore stationary whereas
the other two decay in time. Provided no photon leaks out of the cavity
(which is why conditional dynamics is necessary), a {\em pure} entangled
state between the two
atoms results. From the experimental point of view, this proposal is feasible
 with presently available technology.

The paper is organised as follows. In Section II we describe the system of
interest. This consists of two trapped atoms inside an optical resonator.
Certain aspects of the dynamics of this system when driven by an external
field have been addressed for instance in the context of the two-atom microlaser
\cite{klaus}. The coherence properties of the fluorescence from close lying
atoms in an optical cavity have been considered recently using the quantum
jump approach \cite{carlm}. Our proposal provides a new probabilistic scheme
\cite{other}
for generating an entangled state of the two atoms. This will require an
initial preparation, which involves the selective excitation of
one of the atoms and the continuous monitoring of photons leaking out of the cavity.
The time evolution under the condition of no-photon detection is discussed
in section III. We will show that the quantum jump approach provides a
suitable theoretical framework for analysing the dynamics in a simple and
intuitive way. The fidelity with respect to a
maximally entangled state and the relative entropy of entanglement
of the final atomic state
will be evaluated in section IV.

\section{Description of the physical system.}
Our system consists of two two-level ions confined in a linear trap which has
been surrounded by a leaky optical cavity. We will refer to atom $a$ and atom
$b$ when the context requires us to differentiate them, but otherwise they are
supposed to be identical. We denote by $|0\rangle_i$ and $|1\rangle_i$ the
atomic ground and excited states and with $2\Gamma$ $(\Gamma=\Gamma_a=\Gamma_b)$
the spontaneous emission rate from the upper level. We assume that the distance
between the atoms is much larger than an optical wavelength and that therefore
dipole-dipole interactions can be neglected \cite{prospect}. In addition, this
requirement allows us to assume that each atom can be individually addressed
with laser light. The cavity mode is assumed to be resonant with the atomic
transition frequency and we will denote by $\kappa$ the cavity decay rate.
For the sake of generality we allow the coupling between each atom and the
cavity mode, $g_i$, to be different.\footnote{
A symmetric location of the atoms with respect to the centre of the trap
suffices to make $g_a=g_b$. However, experimentally this may well be hard
to achieve.}
The relaxation of the ion-cavity system can take place through two different
channels, at rates $\kappa$ (cavity decay) and $\Gamma$ (spontaneous decay).

\begin{figure}[htb]
\begin{center}
\epsfxsize7.0cm \centerline{\hspace*{0.cm}\epsfbox{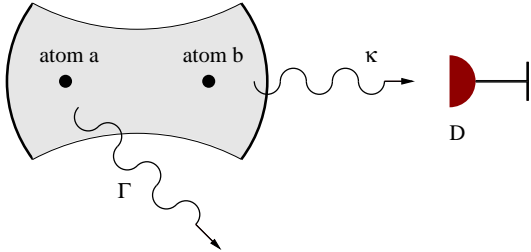} }
\label{fig1}
\end{center}
\caption{\small  Experimental set-up. The system consists of two two-level
atoms placed inside a leaky cavity. The decay rate $\Gamma$ describes the
spontaneous emission of the atoms, while the rate $\kappa$ refers to photons
leaking through the cavity mirrors. The latter can be monitored by the detector
$D$.}
\end{figure}

In what follows we will assume that the coupling constants and the decay rates are such that
\begin{equation}
g_i ,~ \kappa \gg \Gamma.
\end{equation}
The experimental setup is  depicted in Figure 1. Note the presence of a single
photon detector $D$ in our scheme. This set up will allow us to monitor the
decay of the system through the {\em fast} channel, i.e. photons leaking
through the cavity mirrors. On the other hand, spontaneously emitted photons
from the slow decay channel in the regime of Eq. (1), will not be detected.
The initial state of the system is of the form
\begin{equation}
|0\rangle \otimes |0\rangle_a \otimes |0\rangle_b \equiv |000\rangle ,
\end{equation}
where the first index refers to the cavity field state. Applying now a
$\pi$-pulse to atom $a$,
we introduce an excitation into the system and the initial conditions for our
scheme will be given by
the composite state
\begin{equation} \label{initial}
|\psi_0\rangle = |0\rangle \otimes |1\rangle_a \otimes |0\rangle_b \equiv |010\rangle.
\end{equation}
In the following we will use Eq. (\ref{initial}) as the basis for all the
following discussions. It is important to emphasise that our scheme only
requires the atoms to be cooled to the Lamb-Dicke limit, i.e. each atom is
localised within one wavelength of the emitted light. But no further cooling
to the motional ground state is necessary. This notably simplifies the
experimental realisability of the proposal.

\section{The Atom-Cavity system without decay.}
In order to illustrate the main idea underlying this proposal, let us
ignore for the moment any relaxation process. The unitary time evolution
of the system will then be governed by the Hamiltonian
\begin{eqnarray}
H &=& \sum_{i=a,b}\hbar \omega_i |1\rangle_{ii} \langle1|+ \hbar \nu
b^{\dagger}b \nonumber \\
& & + i \hbar  \sum_{i=a,b} \left( g_i b |1\rangle_{ii} \langle 0| -  {\rm h.c.}
\right),
\end{eqnarray}
where $b$ and $b^{\dagger}$ denote the annihilation and creation operators
for the single mode cavity field. The fourth term in this expression is the
familiar Jaynes-Cummings (JC) interaction between each atomic system and the
cavity mode. Moving to an interaction picture with respect to the unperturbed
Hamiltonian
\begin{equation}
H_0=\sum_{i=a,b}\hbar \omega_i |1\rangle_{ii} \langle 1|+ \hbar \nu b^{\dagger}b \end{equation}
and assuming exact resonance between the cavity mode and the atomic transition, $\nu=\omega_i$, we find
\begin{equation}
H_I = i \hbar \sum_{i=a,b} \left( g_i b |1\rangle_{ii} \langle 0| -  {\rm h.c.} \right)
\end{equation}
where the coupling constants $g_i$ have been taken to be real. In the basis
$\cal B$$= (|100\rangle,|010\rangle,|001\rangle)$, the interaction picture
Hamiltonian reads
\begin{equation}
    H_I = \frac{\hbar}{i} \left(
    \begin{array}{ccc}
     0 & g_a & g_b\\
     -g_a & 0 & 0  \\
     -g_b & 0 & 0
    \end{array}\right) .
\end{equation}
It is easy to check that the eigenvalues associated with this operator are given by
\begin{eqnarray}
\lambda_0 &=& 0  \\
\lambda_{1,2} &=& \pm \hbar \, \sqrt{g_a^2+g_b^2}
\end{eqnarray}
with corresponding eigenvectors
\begin{eqnarray}
|\lambda_0\rangle  &=&
\frac{1}{\sqrt{g_a^2+g_b^2}} \, (g_a|001\rangle-g_b|010\rangle) \\ \nonumber
|\lambda_{1,2}\rangle &=&
\frac{1}{\sqrt{2}} \left( |100\rangle \pm
\frac{i}{\sqrt{g_a^2+g_b^2}} \, (g_b|001\rangle+g_a|010\rangle) \right).
\end{eqnarray}
Note that when $g_a=g_b$, the solution $|\lambda_0\rangle$ is a tensor product
of the cavity field
in the vacuum state and the maximally entangled atomic state
\begin{equation}
|\phi^{-}\rangle = \frac{1}{\sqrt{2}}(|01\rangle-|10 \rangle).
\end{equation}
To prepare an entangled state of the atoms one now needs a mechanism that
destroys the population of the cavity mode. One possibility is to use a
leaking cavity and to detect all photons coming through the cavity mirrors.
If a photon is detected the system is in the ground state $|000\rangle$.
Then the experiment has to be repeated. But if not, the systems goes over
into a state which cannot decay. Therefore the atoms should end up in
state $|\lambda_0 \rangle$, the entangled state, where the cavity mode
is not populated.

Using the quantum jump approach we will see that the dynamics under the
condition that no photon has been detected outside the cavity is governed
by an effective Hamiltonian whose solutions keep track of the structure
illustrated above. More precisely, for sufficiently large times the state
of the system will be a tensor product of the cavity field in the vacuum
state and an entangled state of the two atoms.

\section{The Atom-Cavity system including decay.}
Let us consider now the experimental situation depicted in Figure 1, in
which the decay of the cavity field is monitored by means of the detector
D. For the moment we will assume that the detector has $100 \%$ efficiency,
but later this constraint will be relaxed. The time evolution is now
governed by the Hamiltonian
\begin{eqnarray}
H & = & \sum_{i=a,b}\hbar \omega_i |1\rangle_{ii} \langle 1|+ \hbar \nu b^{\dagger}b +
\sum_{{\bf k}\lambda} \hbar \omega_{{\bf k}\lambda}
a_{{\bf k}\lambda}^{\dagger}a_{{\bf k}\lambda} \nonumber \\
& & + i \hbar \sum_{i=a,b} \left( \, g_i b |1\rangle_{ii} \langle 0| - {\rm h.c.} \, \right) \nonumber \\
& & + i \hbar \sum_{i=a,b}\sum_{{\bf k}\lambda} \left( \, g_{{\bf k}\lambda} \, a_{{\bf k}\lambda} |1\rangle_{ii} \langle 0|
e^{i (\omega_i-\omega_{{\bf k}\lambda})t} - {\rm h.c.} \, \right) \nonumber \\
& & + i \hbar \sum_{{\bf k}\lambda} \left( \, s_{{\bf k}\lambda} \, a_{{\bf k}\lambda}b^{\dagger}
e^{i (\nu-\omega_{{\bf k}\lambda})t}
- {\rm h.c.} \, \right),
\end{eqnarray}
where $a_{{\bf k}\lambda}^{\dagger}$ and $a_{{\bf k}\lambda}$ denote the
free radiation field creation and annihilation operators of a photon in
the mode $({\bf k},\lambda)$. The two
remaining terms including the coupling constants $g_{{\bf k}\lambda}$
and $s_{{\bf k}\lambda}$ describe, respectively, the coupling of the
atoms and the cavity mode to the free radiation field. The initial
state of the system , $|\psi_0 \rangle$, is given by Eq. (\ref{initial}).
At a time t, and provided that no photon leaking through the cavity
mirrors has been detected, the state of the system can be described
in terms of a density operator of the form
\begin{eqnarray}
\rho(t,\psi_0) &=& \Big( P_0(t,\psi_0) \, |\hat{\psi}_{coh}(t)\rangle \langle \hat{\psi}_{coh}(t)| \nonumber \\
& & + P_{spon}(t,\psi_0) \, |000\rangle \langle 000| \Big)/{\rm tr}\,( \cdot ).
\end{eqnarray}
Here $P_0(t,\psi_0)$ is the probability for no photon emission, where
neither the cavity field nor the atoms have decayed until $t$, and
$|\hat{\psi}_{coh}(t)\rangle$ denotes the normalised state resulting
from the coherent evolution in this case.  Later we will also use
the notation $|\psi_{coh} \rangle$ for the unnormalised state.
The second term of the mixture takes into account that
spontaneously emitted photons are not observed. If an atom emits a
spontaneous photon, then the state of the atom-cavity system is reduced to the
state $|000 \rangle$. Our main task consists of evaluating the explicit
form of the state $|\hat{\psi}_{coh}(t)\rangle$, of $P_0(t,\psi_0)$ and
the probability $P_{spon}(t,\psi_0)$ for spontaneously decay in $(0,t)$.
The quantum jump approach (also called the quantum trajectories method)
\cite{qj1,qj2,qj3} (See \cite{review} for a recent review) provides a
suitable theoretical framework for this analysis.

\subsection{Derivation of the conditional time evolution.}
Let us consider an idealised situation where both the photons leaking
through the cavity and the spontaneously emitted photons could be
detected. In the derivation of the quantum jump approach one envisages
an equally spaced sequence of gedanken photon measurements at times
$t_1, t_2,...,t_{n-1},t_n$,
such that $t_i-t_{i-1}=\Delta t$. According to the projection postulate,
the sub ensemble for which no photon has been detected until time $t_n$ is
described by the (unnormalised) state vector
\begin{eqnarray}
|\psi_{coh}(t)\rangle &=& \proy_0 U(t_n,t_{n-1}) \proy_0 ...\proy_0 U(t_1,t_0)|0_{ph}\rangle |\psi(t_0)\rangle \nonumber \\
& & \equiv |0_{ph}\rangle U_{cond}(t_n,t_0)|\psi(t_0)\rangle ,
\end{eqnarray}
where we have defined the projector
\begin{equation}
\proy_0 = |0_{ph}\rangle I\!\!I_A \langle 0_{ph}|
\end{equation}
and $I\!\!I_A$ denotes the identity over the atomic variables.
Therefore, the operator $U_{cond}(t_n,t_0)$ describes the time
evolution of the system under the condition that no photon has
been detected. Using our previous notation, the state of the
system at a time $t_n$ will be given by $U_{cond}(t_n,t_0)|\psi(t_0)\rangle$
when the system has not relaxed through either the fast or the slow
channel. Taking into account Eq. (12) and the form of the projector
$\proy_0$, our problem reduces to evaluating expressions of the form
$\langle 0_{ph}|U(t_n,t_{n-1})|0_{ph}\rangle$, which can be
done easily using second order perturbation theory. The calculations can
be simplified moving to an appropriate interaction picture with
respect to the {\em unperturbed} Hamiltonian
\begin{equation}
H_0 = \sum_{i=a,b}\hbar \omega_i |1\rangle_{ii} \langle 1|+ \hbar \nu b^{\dagger}b +
\sum_{{\bf k}\lambda} \hbar \omega_{{\bf k}\lambda} a_{{\bf k}\lambda}^{\dagger}a_{{\bf k}\lambda}.
\end{equation}
In second order perturbation theory one obtains
\begin{eqnarray}
& & \hspace*{-0.8cm} \langle 0_{ph}|U(t_n,t_{n-1})|0_{ph}\rangle \nonumber \\
\hspace*{0.4cm} &=& \id - \frac{1}{\hbar}\int_{t_{n-1}}^{t_n} dt' \langle 0_{ph}|H_I(t')|0_{ph}\rangle \nonumber \\
& & - \frac{1}{\hbar^2} \int_{t_{n-1}}^{t_n} dt' \int_{t_{n-1}}^{t'} dt''
\langle 0_{ph}|H_I(t')H_I(t'')|0_{ph}\rangle,
\label{calculo}
\end{eqnarray}
where the interaction Hamiltonian reads
\begin{eqnarray}
H_I & = &  H_{a-c} + H_{a-f} + H_{c-f} \nonumber \\
& = & i \hbar \sum_{i=a,b} \left(g_i b |1\rangle_{ii} \langle 0| - {\rm h.c.} \right) \nonumber \\
& & + i \hbar \sum_{i=a,b}\sum_{{\bf k}\lambda} \left( g_{{\bf k}\lambda} a_{{\bf k}\lambda} |1\rangle_{ii} \langle 0|
e^{i (\omega_i-\omega_{{\bf k}\lambda})t}- {\rm h.c.} \right) \nonumber \\
    & & + i \hbar \sum_{{\bf k},\lambda} \left( s_{{\bf k}\lambda} a_{{\bf k}\lambda}b^{\dagger} e^{i (\nu-\omega_{{\bf k}\lambda})t} - {\rm h.c.}
\right).
\label{hamil}
\end{eqnarray}
In first order perturbation theory, only the JC-term contributes to
Eq. (17) since both $\langle 0_{ph}|a_{{\bf k}\lambda}|0_{ph}\rangle$ and $\langle 0_{ph}|a_{{\bf k}\lambda}^{\dagger}|0_{ph}\rangle$ are zero. On the
other hand, the second order contribution from the JC term is
quadratic in $g \Delta t$ and can be neglected. A contribution
from the term $H^i_{a-f}$ $(i=a,b)$ appears only in second order
perturbation theory and can be evaluated using the usual Markov
approximation \cite{Cohen-Tannoudji}.
Then one finds
\begin{eqnarray}
& & \hspace*{-0.9cm} - \frac{1}{\hbar^2} \int_{t_{n-1}}^{t_n} dt' \int_{t_{n-1}}^{t'} dt''
\langle 0_{ph}|H_{a-f}(t')H_{a-f}(t'')|0_{ph}\rangle \nonumber \\
\hspace*{0.2cm} &=& - \Gamma_i |1\rangle_{ii} \langle 1| \, \Delta t,
\end{eqnarray}
where
\begin{equation}
\Gamma_i = \frac{e^2}{6 \pi \epsilon_0 \hbar c^3} d^2 \omega_i^3.
\end{equation}
Similarly, one can show that the term $H_{c-f}$ yields a formally
analogous contribution, now replacing the atomic decay rate by the
cavity decay rate $\kappa$. The form of the conditional Hamiltonian
is now easily inferred, taking into account that
\begin{eqnarray}
& &  \hspace*{-0.9cm} \prod_{i=1}^n \langle 0_{ph}|U(t_n,t_{n-1}|0_{ph}\rangle \nonumber \\
\hspace*{0.4cm} &=& U_{cond} (t_n,0)  = {\cal T} \, exp \left( -{i \over \hbar}
\int_0^{t_n} dt' H_{cond} (t') \right)
\end{eqnarray}
where ${\cal T}$ indicates a time ordered expression. We find
\begin{equation} \label{M}
    H_{cond} = \frac{\hbar}{i} \left(
    \begin{array}{ccc}
     \kappa & g_a & g_b\\
     -g_a & \Gamma & 0  \\
     -g_b & 0 & \Gamma
    \end{array}\right) \equiv \frac{\hbar}{i} M
\end{equation}
in the basis $\cal B$$=(|100\rangle ,|010\rangle ,|001\rangle)$. The
corresponding eigenvalues of $M$ are given by
\begin{eqnarray}
\lambda_0 &=& \Gamma  ; \, \\
\lambda_{1,2} &=& (\kappa + \Gamma \pm iS)/2.
\end{eqnarray}
with $ S = \sqrt{4(g_a^2+g_b^2)-(\kappa-\Gamma)^2}$.
The eigenvector of the smallest eigenvalue is the same entangled state as in Eq. (10), i. e.,
\begin{eqnarray}
|\lambda_0\rangle  &=& \frac{1}{\sqrt{g_a^2+g_b^2}} \, (g_a|001\rangle-g_b|010\rangle) .
\end{eqnarray}
$M$ has three normalised
eigenvectors $|\lambda_i \rangle$, which are in general not
orthogonal. The reciprocal vectors $\langle \lambda^i|$  are
defined by $\langle \lambda^i|\lambda_j \rangle=\delta_{ij}$.
Then one can write $M=\sum_i \lambda_i |\lambda_i \rangle \langle \lambda^i|$.
For the conditional time evolution operator one has the representation
\begin{equation}
U_{cond}(t,0)=e^{-Mt}=\sum_{i=1}^3 e^{-\lambda_it} \,
|\lambda_i \rangle \langle \lambda^i| .
\end{equation}
Therefore, provided that no photon has been detected during the time
interval $[0,t]$ and $t$ satisfies
\begin{equation}
\Gamma^{-1} \gg t \gg \kappa^{-1}
\label{eq25}
\end{equation}
the exponentials $\exp(-\lambda_{1/2}t)$ can be neglected while
$\exp(-\lambda_0t)$ is still close to unity and the system will
be in the state
\begin{eqnarray}
| \hat{\psi}_{coh}(t) \rangle &=& U_{cond}(t,0) |\psi_0 \rangle \nonumber \\
&=& e^{-\lambda_0 t} \,
|\lambda_0 \rangle \langle \lambda^0|\psi_0 \rangle /\| \cdot \|
= |\lambda_0 \rangle.
\end{eqnarray}
This state factorises as a tensor product between the cavity field in the
 vacuum state and an entangled state of the two atoms.

More precisely, the conditional time evolution operator $U_{cond}$ can be calculated as
\begin{eqnarray}
e^{-Mt} &=&  \frac{(M-\lambda_1)(M-\lambda_2)}{(\lambda_0-\lambda_1)(\lambda_0-\lambda_2)}
\, e^{-\lambda_0t} \nonumber \\
& & + ({\rm cyclic~permutations}),
\end{eqnarray}
which can easily be verified by application to the eigenvectors \cite{gant}.
Applying this operator to our initial state, Eq. (\ref{initial}), we obtain
\end{multicols}
\noindent\rule{0.5\textwidth}{0.4pt}\rule{0.4pt}{\baselineskip}
\begin{eqnarray}
|\hat{\psi}_{coh}(t) \rangle &=& \frac{1}{g_a^2+g_b^2} \left[
g_b \, e^{-\Gamma t} \left(
    \begin{array}{r}
     0\\
     g_b\\
     -g_a \end{array}\right)
+ g_a \, e^{-\frac{1}{2}(\kappa+\Gamma)t} \, \left\{ \left(
    \begin{array}{r}
     0\\
     g_a\\
     g_b \end{array}\right) \cos(S t/2)
+ \frac{1}{S} \left(
    \begin{array}{r}
     -2 (g_a^2+g_b^2)\\
     g_a (\kappa-\Gamma)\\
     g_b (\kappa-\Gamma) \end{array}\right) \sin(S t/2) \right\} \right].
    \label{coh}
\end{eqnarray}
\hspace*{\fill}\rule[0.4pt]{0.4pt}{\baselineskip}%
\rule[\baselineskip]{0.5\textwidth}{0.4pt}
\begin{multicols}{2}

The probability amplitudes for the three basis states are plotted in
Figure 2.
\end{multicols}
\begin{minipage}{6.54truein}
\begin{figure}[htb]
\begin{center}
\input{fig2.tex}
\label{fig2}
\end{center}
\caption{\small The time dependence of the probability amplitudes for the basis states $|100\rangle$, $|010\rangle$ and $|001\rangle$ under the conditional time evolution that no photon has been detected at all. We have chosen $g_a=g_b=g=\kappa$ and $\Gamma = 10^{-3} g$. After a short time the cavity mode is decayed and the atoms have reached the pure entangled atomic state.}

\end{figure}
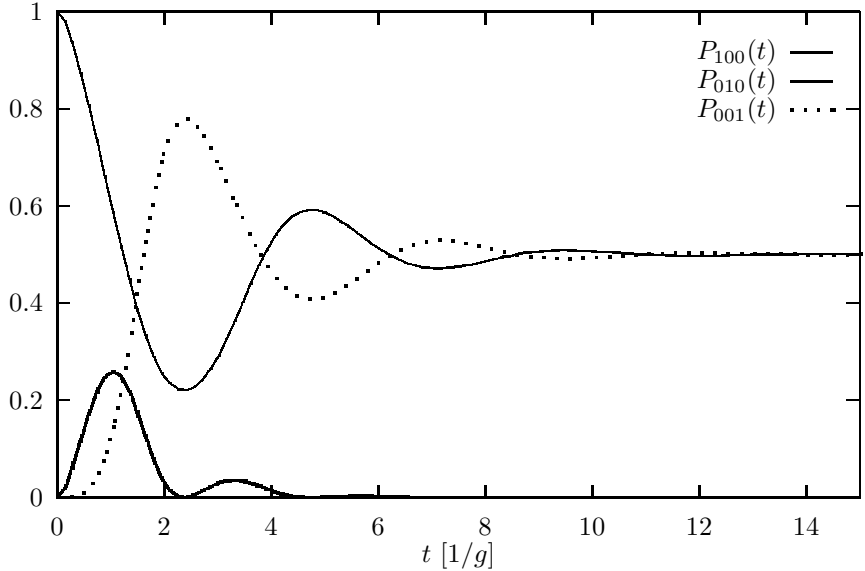
\end{minipage}
\begin{multicols}{2}
As expected, in a time scale such that $\Gamma^{-1} \gg t \gg \kappa^{-1}$,
the contribution from terms multiplied by a damping factor proportional to
the sum $\kappa + \Gamma$ becomes negligible and the conditional state
vector is a two-particle entangled state correlated with the cavity field in the vacuum state $|\lambda_0 \rangle$.

\subsection{Calculation of the detection probabilities.}
After the derivation of the conditional time evolution we are now in a
position to calculate the probabilities for photon emissions. We first
calculate the probability that there is no decay at all, neither spontaneous
emissions by the atoms nor photons leaking out of the cavity. Subsequently we will
derive the probability for (a) having a spontaneous decay from the atoms and
(b) for having photon emission from the cavity.

The probability to have no photon emission (neither spontaneously emitted
nor leaking through the cavity mirrors) until time $t$ is given by the
norm squared of Eq. (\ref{coh}), i.e.
\begin{equation}
P_0 (t,\psi_0) = \parallel U_{cond}(t,0) | \psi_0 \rangle \parallel^2 .
\end{equation}
This general expression can be simplified considerably for large times $t$.
The probability to detect no photon until time
$t$ with $t \gg \kappa^{-1}$ is equal to
\begin{equation}
    P_0(t,\psi_0) = \frac{g_b^2}{g_a^2+g_b^2} \, e^{-2 \Gamma t} .
\end{equation}

In our experimental set up (see Figure 1) only photons
leaking through the cavity mirrors are monitored and, as we have
pointed out, the state of the system
will be the mixture given by Eq. (13). The quantum jump approach
\cite{qj2,qj3,review}
provides a transparent way to evaluate the weight of the component
$| 000 \rangle $, i.e. the probability for a spontaneous emission from an atom.

Let us denote by $t'$ an intermediate time within the interval $[0,t]$.
The probability $P$ of having an {\em emission} at any time in that
interval will be given by
\begin{eqnarray} \label{qj}
    P &=& \int_{0}^t dt' \, w_1(t',\psi_0) ,
\end{eqnarray}
where $w_1(t',\psi_0)$ denotes the probability density for the first photon at time $t'$ for the given initial state $|\psi_0\rangle$ \cite{PLeniootro,kim}.
Since $w_1(t',\psi_0) \, dt$ equals $P_0(t',\psi_0)-P_0(t'+dt',\psi_0)$
one has
\begin{eqnarray}
w_1(t',\psi_0) &=& - \frac{d}{dt'} P_0(t',\psi_0) \nonumber \\
&=& \langle \psi_0 |e^{-M^{\dagger} t' } (M+M^\dagger) e^{-M t'}| \psi_0 \rangle.
\end{eqnarray}
Taking into account the explicit form of $M$ in Eq. (\ref{M}), we find
\begin{eqnarray}
w_1(t',\psi_0) &=& 2 \kappa \, |\langle 100| U_{cond}(t',0)|\psi_0 \rangle|^2
\nonumber \\
& & + 2 \Gamma \, \left( \, |\langle 010|U_{cond}(t',0)| \psi_0 \rangle|^2
\right. \nonumber \\
& & + \left. |\langle 001| U_{cond}(t',0)| \psi_0 \rangle|^2 \, \right).
\end{eqnarray}
As expected, both relaxation channels contribute separately to the
decay rate $w_1$. Setting $t'$ equal to 0 one finds that the
probability density for a photon leaking through the cavity
mirrors is given by the population of the state $|100 \rangle$
multiplied by the cavity decay rate. Similarly the probability
for spontaneous emission is determined by the population of the
states $|010 \rangle$ and $|001 \rangle$.

In our case we are only interested in the contribution to $P$ in
Eq. (\ref{qj}) coming from spontaneously emitted photons. Using
Eq. (\ref{qj}) one finds
\begin{eqnarray}
P_{spon}(t,\psi_0) &=& 2 \Gamma \, \int_0^t dt' \left( |\langle 010 | U_{cond}(t',0)|\psi_0 \rangle |^2 \right. \nonumber \\
& & + \left. |\langle 001 | U_{cond}(t',0)|\psi_0 \rangle |^2 \right).
\end{eqnarray}
However from the point of view of simplifying the calculations
it is easier to evaluate the probability of cavity decay. In a
similar way one obtains
\begin{equation}
P_{cav}(t,\psi_0) =  2 \kappa \, \int_0^t dt'
|\langle 010 | U_{cond}(t',0)|\psi_0 \rangle |^2.
\end{equation}
Taking into account the results of the previous section for the
unnormalised state $|\psi_{coh} \rangle$, we can write
\begin{eqnarray}
P_{cav}(t,\psi_0) &=& \frac{\kappa g_a^2}{(\kappa+\Gamma)(g_a^2+g_b^2+\kappa\Gamma)} \, \nonumber \\
& & \hspace*{-0.7cm} \Big[ 1 -\frac{e^{-(\kappa+\Gamma)t}}{S^2} \, \Big(4 (g_a^2+g_b^2+\kappa\Gamma) \nonumber \\
& & \hspace*{-0.7cm} +(\kappa+\Gamma)\left(S \sin(S t)-(\kappa+\Gamma)
\cos(S t) \right) \Big) \Big]
\end{eqnarray}
and calculate $P_{spon}$ as the difference between unity and the
sum $P_0+P_{cav}$.

\section{Fidelity and entanglement in the asymptotic regime.}
In the previous section we have derived the exact analytical expressions
for the no-decay probabilities. In this section we will now discuss these
exact expressions in the {\em asymptotic} regime, i.e. for times longer
than the cavity lifetime. Finally, we will characterise the quality of
the entanglement generation by cavity loss in two ways. We will calculate
the fidelity with respect to the maximally entangled state
$ | \phi^{-} \rangle $ and we will calculate explicitly a measure of
entanglement (the relative entropy of entanglement \cite{matpra}) for
the state of the system.

In the Figure 3 we plot the probability $P_{cav}(t,\psi_0)$ that a photon has
leaked out of the cavity.
\end{multicols}
\begin{minipage}{6.54truein}
\begin{figure}[htb]
\begin{center}
\input{fig3.tex}
\label{fig3}
\end{center}
\caption{\small The probability for the photon leaking through the cavity mirrors in the time
interval $[0,t]$. We have chosen $g_a=g_b=g=\kappa$ and $\Gamma = 10^{-3} g$. For these
parameters the cavity mode decays with a probability close to 1/2.
After a short time the state inside the cavity is stable.}
\end{figure}
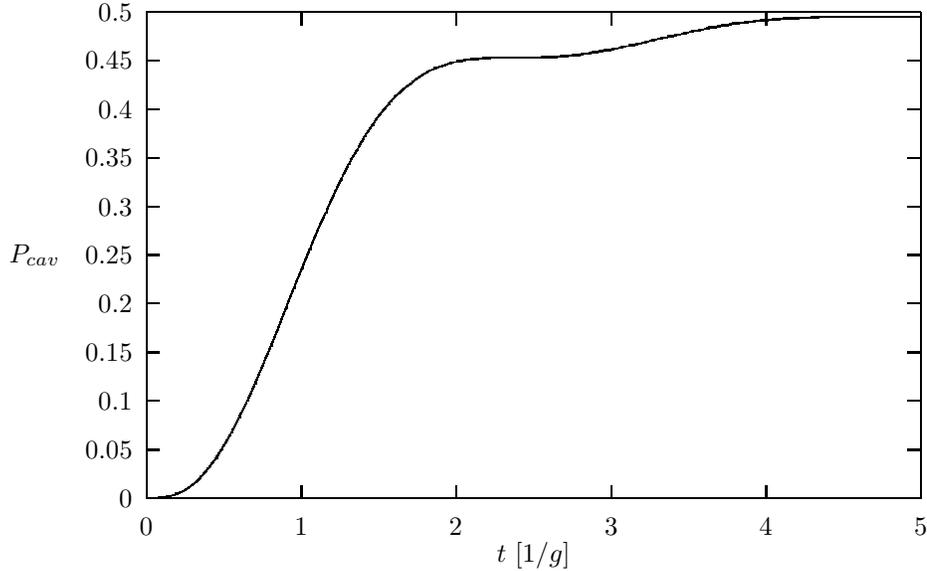
\end{minipage}
\begin{multicols}{2}
As expected, this function saturates at a point close
to $0.5$ when $g_a=g_b$ and $\Gamma$ is small. The reason for this is the overlap
of the initial state $|010\rangle$ with the singlet state $|0\rangle| \phi^{-} \rangle$
is precisely $1/2$. If a photon leaks the cavity, then the atomic state is $|00\rangle$,
i.e. the atomic state is a product state. If no photon leaks out of the cavity then the
atoms are in an entangled state. Therefore the scheme presented here succeeds in $50\%$
of the cases. In the asymptotic regime we can write
\begin{eqnarray}
P_{spon}(t,\psi_0)
&=& 1 - \frac{g_b^2}{g_a^2+g_b^2} \, e^{-2 \Gamma t} \nonumber \\
& & - \frac{g_a^2 \, \kappa}{(\Gamma+\kappa)(g_a^2+g_b^2+\Gamma\kappa)}.
\end{eqnarray}
Using the expressions for $|\psi_{coh}(t)\rangle$ and $P_{spon}(t,\psi_0)$ we can now
calculate the state of the atoms at time $t$. This expression can then
be used to evaluate the fidelity with respect to the maximally entangled state
$| \phi^{-} \rangle$ of Eq. (11). This result has been represented in Figure 4.
\end{multicols}
\begin{minipage}{6.54truein}
\begin{figure}[htb]
\begin{center}
\input{fig4.tex}
\label{fig4}
\end{center}
\caption{\small Fidelity of the final atomic state with respect to the singlet state
in the asymptotic limit, where $t$ is large compared with $\kappa^{-1}$.
The dotted line corresponds to the case of a detector with finite efficiency $\eta$
(here $\eta=0.8$). For small times the fidelity of the atomic state with respect to
the singlet state is high, even for a counter efficiency $\eta=0.8$). For larger times
the fidelity decreases exponentially because of a spontaneously emitted photon.}

\end{figure}
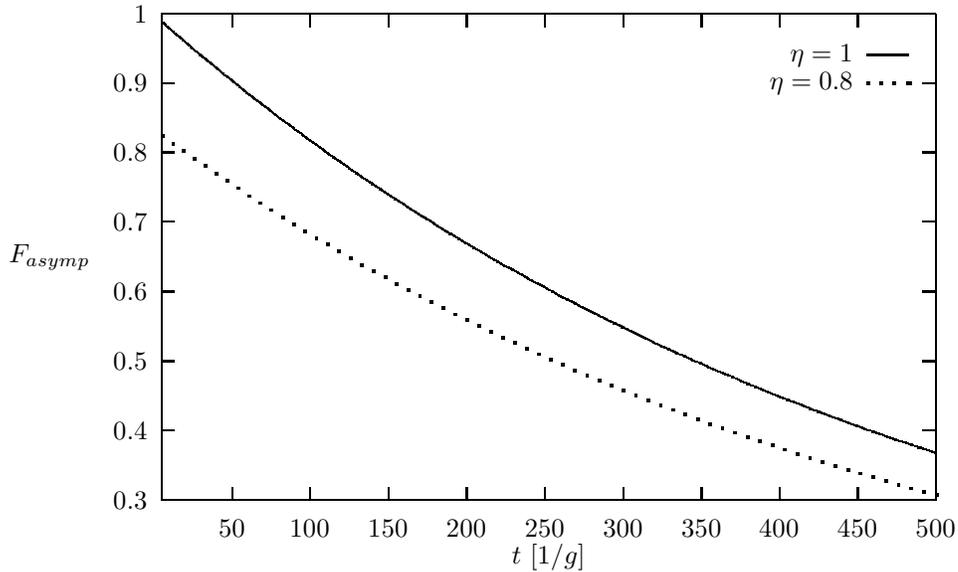
\end{minipage}
\begin{multicols}{2}

We observe that for short times $t$ satisfying Eq. (\ref{eq25}), the fidelity is   almost unity. For times comparable or larger than $\Gamma^{-1}$ the fidelity falls off exponentially. For our proposal only the region with small $t$ is relevant, so that the exponential decay of the fidelity for larger $t$ does not limit the efficiency of our scheme. In Figure 4 we also plotted the fidelity for imperfect counter efficiency (in this figure is $\eta=0.8$). We observe that the fidelity is still high.

When dealing with entangled states it is interesting to know the amount
of entanglement that is contained in a state. Especially for mixed states
this is not directly related to the fidelity of the state. However, there
exist quantitative entanglement measures for mixed states. In the following
we will calculate the relative entropy of entanglement for the states
generated by our scheme.
Due to the special form of the density operator $\rho$ of the two atoms
\begin{eqnarray}
\rho &=& \frac{1}{P_0(t,\psi_0)+P_{spon}(t,\psi_0)} \nonumber \\
& & \left( P_0(t,\psi_0) \, | \phi^{-} \rangle \langle \phi^{-}|
+ P_{spon}(t,\psi_0) \, | 00 \rangle \langle 00| \right) \label{rho}, \label{state}
\end{eqnarray}
it is possible to compute the relative entropy of entanglement of the
final state \cite{matpra} analytically. It is given by
\begin{equation}
E(\rho) = (\lambda-2) \log_2(1-\lambda/2) + (1-\lambda) \log_2(1-\lambda)
\end{equation}
where $\lambda = P_0/(P_0+P_{spon})$.
We have plotted this result in Figure 5 for perfect counter efficiency.
\end{multicols}
\begin{minipage}{6.54truein}
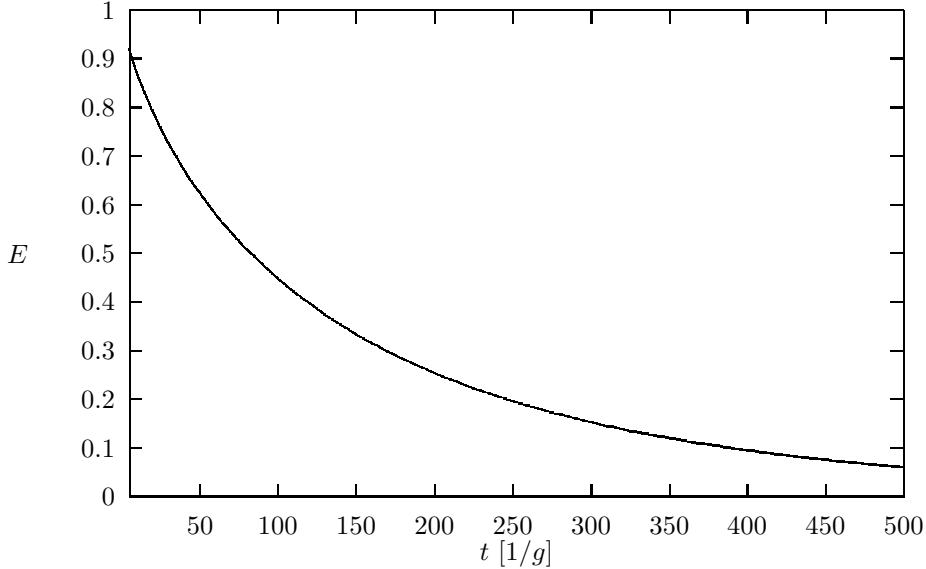
\begin{figure}[htb]
\begin{center}
\input{fig5.tex}
\label{fig5}
\end{center}
\caption{\small  Relative entropy of entanglement for the final mixed state in the asymptotic limit,
where $t$ is large compared with $\kappa^{-1}$. As before we have taken $g_a=g_b=g=\kappa$
and $\Gamma = 10^{-3} g$. As long, as the entangled state of the atoms does not decay
spontaneously, the entropy $E$ is high.}

\end{figure}
\end{minipage}
\begin{multicols}{2}
For short times (which are nevertheless longer than the cavity lifetime)
the amount of entanglement is high while it falls off exponentially for larger times.
It should be noted that the state Eq. (\ref{state}) contains entanglement
for arbitrary counter efficiencies and spontaneous decay rates of the atoms.
Therefore our scheme is not limited by these experimental imperfections.

The fidelity of the mixed state $\rho$ can be determined
experimentally using the technique recently developed by the NIST
group in Colorado \cite{quentin}. Both the diagonal elements and
the relevant off-diagonal coherences of mixed states of the form
of Eq. (\ref{rho}) can be measured by this method. Note that our
approach allows us to incorporate easily a non-unit efficiency for
the photo detectors. All we have to do is to modify the weight of
the component $| 000 \rangle$ to account for the fact that there
is a finite probability $\eta$ that the photo detector has not
triggered in spite of the fact that leaking has occurred. The
weight $P_{spon}$ is then replaced by $P_{spon} + (1-\eta)P_{cav}$
\cite{PLeniootro}. The effect of non-ideal detectors on the
fidelity of the state is illustrated by the dotted line of Figure
3. For a counter efficiency of $80\%$ the fidelity of the atomic
state with respect to the singlet state is still high. Note that
the effect of a nonperfect counter or spontaneous emission can be
corrected using the following idea. A nonperfect counter or
spontaneous emission lead to a $|000\rangle$ contribution in the
density operator; see Eq. (40). If we irradiate a system in a
state $|000\rangle$ by a laser, cavity photons will be excited
which will eventually leak out of the cavity mirror where they
will be detected. The singlet contribution to the density operator
remains invariant under the same procedure. In the state Eq. (40)
only the $|000\rangle$ contribution will lead to the detection of
a cavity photon. If we detect such a photon, the state of the
system is projected to the state $|000\rangle$. If we fail to
detect a photon, then even for imperfect counters, we will end up
in a state that has a higher proportion of the singlet state. Few
repetitions of this procedure reduce the $|000\rangle$
contribution in the density operator of the atoms to very low
values. Therefore, we conclude that our scheme is not overly
sensitive to counter efficiency.

\section{Conclusions}
We have described an experimental situation where entanglement between
two atomic systems can be induced via continuous observation of the cavity
loss. This proposal allows us to illustrate the effects of conditional time
evolution and the power of the quantum jump approach as an analytical
tool. From the experimental point of view
the proposal has a number of advantages that should make its experimental
realization possible with existing experimental methods.
\begin{enumerate}
\item There exist open ion traps that allow to implement a sufficiently
small cavity. This will allow us to achieve high coupling constants between
atoms and cavity.
\item The condition given by Eq. (1) are experimentally achievable as
we do not require
the strong coupling regime.
\item The atoms only need to be cooled to the Lamb-Dicke limit. In present
ion trap implementations of entanglement manipulations the cooling to the
motional ground state of the ions is required. For more than a single ion
this can, at present, only be achieved with a finite precision and currently represents
a strong limit to the achievable fidelity of the state of the entangled atoms
\cite{quentin}.
\item The detection efficiency varies with the wavelength but it can be up to
$90 \%$. Although the amount of entanglement in the atomic state decreases with
decreasing counter efficiency it never vanishes (see also Figure 4).
\end{enumerate}
In addition, the initial preparation requires only a single laser pulse to
excite selectively one of the atoms. Therefore, the experiment proposed here
does seem feasible with presently available technology. \\

\noindent {\bf Acknowledgements} \\ \noindent We thank J.
Steinbach and D. M. Segal for useful comments on the subject of
this paper. Part of this work has been done during the 1998
Quantum Optics and Quantum Computing Workshop at the Benasque
Center of Physics. This work was supported by the European
Community, the UK Engineering and Physical Sciences Research
Council and by a Feodor Lynen grant of the Alexander von Humboldt
Foundation and by the Leverhulme Trust.

\end{multicols}

\end{document}

%% file: fig2.tex
\setlength{\unitlength}{0.240900pt}
\ifx\plotpoint\undefined\newsavebox{\plotpoint}\fi
\sbox{\plotpoint}{\rule[-0.200pt]{0.400pt}{0.400pt}}%
\begin{picture}(1500,900)(0,0)
\font\gnuplot=cmr10 at 10pt
\gnuplot
\sbox{\plotpoint}{\rule[-0.200pt]{0.400pt}{0.400pt}}%
\put(176.0,113.0){\rule[-0.200pt]{303.534pt}{0.400pt}}
\put(176.0,113.0){\rule[-0.200pt]{0.400pt}{184.048pt}}
\put(176.0,113.0){\rule[-0.200pt]{4.818pt}{0.400pt}}
\put(154,113){\makebox(0,0)[r]{0}}
\put(1416.0,113.0){\rule[-0.200pt]{4.818pt}{0.400pt}}
\put(176.0,266.0){\rule[-0.200pt]{4.818pt}{0.400pt}}
\put(154,266){\makebox(0,0)[r]{0.2}}
\put(1416.0,266.0){\rule[-0.200pt]{4.818pt}{0.400pt}}
\put(176.0,419.0){\rule[-0.200pt]{4.818pt}{0.400pt}}
\put(154,419){\makebox(0,0)[r]{0.4}}
\put(1416.0,419.0){\rule[-0.200pt]{4.818pt}{0.400pt}}
\put(176.0,571.0){\rule[-0.200pt]{4.818pt}{0.400pt}}
\put(154,571){\makebox(0,0)[r]{0.6}}
\put(1416.0,571.0){\rule[-0.200pt]{4.818pt}{0.400pt}}
\put(176.0,724.0){\rule[-0.200pt]{4.818pt}{0.400pt}}
\put(154,724){\makebox(0,0)[r]{0.8}}
\put(1416.0,724.0){\rule[-0.200pt]{4.818pt}{0.400pt}}
\put(176.0,877.0){\rule[-0.200pt]{4.818pt}{0.400pt}}
\put(154,877){\makebox(0,0)[r]{1}}
\put(1416.0,877.0){\rule[-0.200pt]{4.818pt}{0.400pt}}
\put(176.0,113.0){\rule[-0.200pt]{0.400pt}{4.818pt}}
\put(176,68){\makebox(0,0){0}}
\put(176.0,857.0){\rule[-0.200pt]{0.400pt}{4.818pt}}
\put(344.0,113.0){\rule[-0.200pt]{0.400pt}{4.818pt}}
\put(344,68){\makebox(0,0){2}}
\put(344.0,857.0){\rule[-0.200pt]{0.400pt}{4.818pt}}
\put(512.0,113.0){\rule[-0.200pt]{0.400pt}{4.818pt}}
\put(512,68){\makebox(0,0){4}}
\put(512.0,857.0){\rule[-0.200pt]{0.400pt}{4.818pt}}
\put(680.0,113.0){\rule[-0.200pt]{0.400pt}{4.818pt}}
\put(680,68){\makebox(0,0){6}}
\put(680.0,857.0){\rule[-0.200pt]{0.400pt}{4.818pt}}
\put(848.0,113.0){\rule[-0.200pt]{0.400pt}{4.818pt}}
\put(848,68){\makebox(0,0){8}}
\put(848.0,857.0){\rule[-0.200pt]{0.400pt}{4.818pt}}
\put(1016.0,113.0){\rule[-0.200pt]{0.400pt}{4.818pt}}
\put(1016,68){\makebox(0,0){10}}
\put(1016.0,857.0){\rule[-0.200pt]{0.400pt}{4.818pt}}
\put(1184.0,113.0){\rule[-0.200pt]{0.400pt}{4.818pt}}
\put(1184,68){\makebox(0,0){12}}
\put(1184.0,857.0){\rule[-0.200pt]{0.400pt}{4.818pt}}
\put(1352.0,113.0){\rule[-0.200pt]{0.400pt}{4.818pt}}
\put(1352,68){\makebox(0,0){14}}
\put(1352.0,857.0){\rule[-0.200pt]{0.400pt}{4.818pt}}
\put(176.0,113.0){\rule[-0.200pt]{303.534pt}{0.400pt}}
\put(1436.0,113.0){\rule[-0.200pt]{0.400pt}{184.048pt}}
\put(176.0,877.0){\rule[-0.200pt]{303.534pt}{0.400pt}}
\put(806,23){\makebox(0,0){$t~[1/g]$}}
\put(176.0,113.0){\rule[-0.200pt]{0.400pt}{184.048pt}}
\sbox{\plotpoint}{\rule[-0.400pt]{0.800pt}{0.800pt}}%
\put(1306,812){\makebox(0,0)[r]{$P_{100}(t)$}}
\put(1328.0,812.0){\rule[-0.400pt]{15.899pt}{0.800pt}}
\put(176,113){\usebox{\plotpoint}}
\multiput(177.41,113.00)(0.509,0.574){19}{\rule{0.123pt}{1.123pt}}
\multiput(174.34,113.00)(13.000,12.669){2}{\rule{0.800pt}{0.562pt}}
\multiput(190.41,128.00)(0.511,1.530){17}{\rule{0.123pt}{2.533pt}}
\multiput(187.34,128.00)(12.000,29.742){2}{\rule{0.800pt}{1.267pt}}
\multiput(202.41,163.00)(0.509,1.733){19}{\rule{0.123pt}{2.846pt}}
\multiput(199.34,163.00)(13.000,37.093){2}{\rule{0.800pt}{1.423pt}}
\multiput(215.41,206.00)(0.509,1.650){19}{\rule{0.123pt}{2.723pt}}
\multiput(212.34,206.00)(13.000,35.348){2}{\rule{0.800pt}{1.362pt}}
\multiput(228.41,247.00)(0.509,1.402){19}{\rule{0.123pt}{2.354pt}}
\multiput(225.34,247.00)(13.000,30.114){2}{\rule{0.800pt}{1.177pt}}
\multiput(241.41,282.00)(0.511,0.943){17}{\rule{0.123pt}{1.667pt}}
\multiput(238.34,282.00)(12.000,18.541){2}{\rule{0.800pt}{0.833pt}}
\multiput(252.00,305.40)(1.000,0.526){7}{\rule{1.686pt}{0.127pt}}
\multiput(252.00,302.34)(9.501,7.000){2}{\rule{0.843pt}{0.800pt}}
\multiput(265.00,309.08)(0.737,-0.516){11}{\rule{1.356pt}{0.124pt}}
\multiput(265.00,309.34)(10.186,-9.000){2}{\rule{0.678pt}{0.800pt}}
\multiput(279.41,295.04)(0.509,-0.947){19}{\rule{0.123pt}{1.677pt}}
\multiput(276.34,298.52)(13.000,-20.519){2}{\rule{0.800pt}{0.838pt}}
\multiput(292.41,267.76)(0.511,-1.485){17}{\rule{0.123pt}{2.467pt}}
\multiput(289.34,272.88)(12.000,-28.880){2}{\rule{0.800pt}{1.233pt}}
\multiput(304.41,233.46)(0.509,-1.526){19}{\rule{0.123pt}{2.538pt}}
\multiput(301.34,238.73)(13.000,-32.731){2}{\rule{0.800pt}{1.269pt}}
\multiput(317.41,195.97)(0.509,-1.443){19}{\rule{0.123pt}{2.415pt}}
\multiput(314.34,200.99)(13.000,-30.987){2}{\rule{0.800pt}{1.208pt}}
\multiput(330.41,161.42)(0.511,-1.214){17}{\rule{0.123pt}{2.067pt}}
\multiput(327.34,165.71)(12.000,-23.711){2}{\rule{0.800pt}{1.033pt}}
\multiput(342.41,136.32)(0.509,-0.740){19}{\rule{0.123pt}{1.369pt}}
\multiput(339.34,139.16)(13.000,-16.158){2}{\rule{0.800pt}{0.685pt}}
\multiput(354.00,121.08)(0.737,-0.516){11}{\rule{1.356pt}{0.124pt}}
\multiput(354.00,121.34)(10.186,-9.000){2}{\rule{0.678pt}{0.800pt}}
\put(367,111.84){\rule{3.132pt}{0.800pt}}
\multiput(367.00,112.34)(6.500,-1.000){2}{\rule{1.566pt}{0.800pt}}
\put(380,113.34){\rule{2.600pt}{0.800pt}}
\multiput(380.00,111.34)(6.604,4.000){2}{\rule{1.300pt}{0.800pt}}
\multiput(392.00,118.40)(1.000,0.526){7}{\rule{1.686pt}{0.127pt}}
\multiput(392.00,115.34)(9.501,7.000){2}{\rule{0.843pt}{0.800pt}}
\multiput(405.00,125.39)(1.244,0.536){5}{\rule{1.933pt}{0.129pt}}
\multiput(405.00,122.34)(8.987,6.000){2}{\rule{0.967pt}{0.800pt}}
\multiput(418.00,131.39)(1.244,0.536){5}{\rule{1.933pt}{0.129pt}}
\multiput(418.00,128.34)(8.987,6.000){2}{\rule{0.967pt}{0.800pt}}
\put(431,135.84){\rule{2.891pt}{0.800pt}}
\multiput(431.00,134.34)(6.000,3.000){2}{\rule{1.445pt}{0.800pt}}
\put(443,137.84){\rule{3.132pt}{0.800pt}}
\multiput(443.00,137.34)(6.500,1.000){2}{\rule{1.566pt}{0.800pt}}
\put(456,137.84){\rule{3.132pt}{0.800pt}}
\multiput(456.00,138.34)(6.500,-1.000){2}{\rule{1.566pt}{0.800pt}}
\put(469,135.34){\rule{2.600pt}{0.800pt}}
\multiput(469.00,137.34)(6.604,-4.000){2}{\rule{1.300pt}{0.800pt}}
\put(481,131.34){\rule{2.800pt}{0.800pt}}
\multiput(481.00,133.34)(7.188,-4.000){2}{\rule{1.400pt}{0.800pt}}
\multiput(494.00,129.06)(1.768,-0.560){3}{\rule{2.280pt}{0.135pt}}
\multiput(494.00,129.34)(8.268,-5.000){2}{\rule{1.140pt}{0.800pt}}
\multiput(507.00,124.06)(1.768,-0.560){3}{\rule{2.280pt}{0.135pt}}
\multiput(507.00,124.34)(8.268,-5.000){2}{\rule{1.140pt}{0.800pt}}
\put(520,117.84){\rule{2.891pt}{0.800pt}}
\multiput(520.00,119.34)(6.000,-3.000){2}{\rule{1.445pt}{0.800pt}}
\put(532,114.84){\rule{3.132pt}{0.800pt}}
\multiput(532.00,116.34)(6.500,-3.000){2}{\rule{1.566pt}{0.800pt}}
\put(545,112.84){\rule{3.132pt}{0.800pt}}
\multiput(545.00,113.34)(6.500,-1.000){2}{\rule{1.566pt}{0.800pt}}
\put(558,111.84){\rule{3.132pt}{0.800pt}}
\multiput(558.00,112.34)(6.500,-1.000){2}{\rule{1.566pt}{0.800pt}}
\put(583,111.84){\rule{3.132pt}{0.800pt}}
\multiput(583.00,111.34)(6.500,1.000){2}{\rule{1.566pt}{0.800pt}}
\put(571.0,113.0){\rule[-0.400pt]{2.891pt}{0.800pt}}
\put(609,112.84){\rule{2.891pt}{0.800pt}}
\multiput(609.00,112.34)(6.000,1.000){2}{\rule{1.445pt}{0.800pt}}
\put(596.0,114.0){\rule[-0.400pt]{3.132pt}{0.800pt}}
\put(634,113.84){\rule{3.132pt}{0.800pt}}
\multiput(634.00,113.34)(6.500,1.000){2}{\rule{1.566pt}{0.800pt}}
\put(621.0,115.0){\rule[-0.400pt]{3.132pt}{0.800pt}}
\put(660,113.84){\rule{2.891pt}{0.800pt}}
\multiput(660.00,114.34)(6.000,-1.000){2}{\rule{1.445pt}{0.800pt}}
\put(647.0,116.0){\rule[-0.400pt]{3.132pt}{0.800pt}}
\put(698,112.84){\rule{3.132pt}{0.800pt}}
\multiput(698.00,113.34)(6.500,-1.000){2}{\rule{1.566pt}{0.800pt}}
\put(672.0,115.0){\rule[-0.400pt]{6.263pt}{0.800pt}}
\put(723,111.84){\rule{3.132pt}{0.800pt}}
\multiput(723.00,112.34)(6.500,-1.000){2}{\rule{1.566pt}{0.800pt}}
\put(711.0,114.0){\rule[-0.400pt]{2.891pt}{0.800pt}}
\put(736.0,113.0){\rule[-0.400pt]{168.630pt}{0.800pt}}
\sbox{\plotpoint}{\rule[-0.200pt]{0.400pt}{0.400pt}}%
\put(1306,767){\makebox(0,0)[r]{$P_{010}(t)$}}
\put(1328.0,767.0){\rule[-0.200pt]{15.899pt}{0.400pt}}
\put(176,877){\usebox{\plotpoint}}
\multiput(176.58,874.67)(0.493,-0.576){23}{\rule{0.119pt}{0.562pt}}
\multiput(175.17,875.83)(13.000,-13.834){2}{\rule{0.400pt}{0.281pt}}
\multiput(189.58,856.60)(0.492,-1.530){21}{\rule{0.119pt}{1.300pt}}
\multiput(188.17,859.30)(12.000,-33.302){2}{\rule{0.400pt}{0.650pt}}
\multiput(201.58,819.45)(0.493,-1.884){23}{\rule{0.119pt}{1.577pt}}
\multiput(200.17,822.73)(13.000,-44.727){2}{\rule{0.400pt}{0.788pt}}
\multiput(214.58,770.94)(0.493,-2.043){23}{\rule{0.119pt}{1.700pt}}
\multiput(213.17,774.47)(13.000,-48.472){2}{\rule{0.400pt}{0.850pt}}
\multiput(227.58,718.56)(0.493,-2.162){23}{\rule{0.119pt}{1.792pt}}
\multiput(226.17,722.28)(13.000,-51.280){2}{\rule{0.400pt}{0.896pt}}
\multiput(240.58,662.84)(0.492,-2.392){21}{\rule{0.119pt}{1.967pt}}
\multiput(239.17,666.92)(12.000,-51.918){2}{\rule{0.400pt}{0.983pt}}
\multiput(252.58,607.43)(0.493,-2.201){23}{\rule{0.119pt}{1.823pt}}
\multiput(251.17,611.22)(13.000,-52.216){2}{\rule{0.400pt}{0.912pt}}
\multiput(265.58,551.69)(0.493,-2.122){23}{\rule{0.119pt}{1.762pt}}
\multiput(264.17,555.34)(13.000,-50.344){2}{\rule{0.400pt}{0.881pt}}
\multiput(278.58,497.94)(0.493,-2.043){23}{\rule{0.119pt}{1.700pt}}
\multiput(277.17,501.47)(13.000,-48.472){2}{\rule{0.400pt}{0.850pt}}
\multiput(291.58,446.08)(0.492,-2.004){21}{\rule{0.119pt}{1.667pt}}
\multiput(290.17,449.54)(12.000,-43.541){2}{\rule{0.400pt}{0.833pt}}
\multiput(303.58,400.48)(0.493,-1.567){23}{\rule{0.119pt}{1.331pt}}
\multiput(302.17,403.24)(13.000,-37.238){2}{\rule{0.400pt}{0.665pt}}
\multiput(316.58,361.37)(0.493,-1.290){23}{\rule{0.119pt}{1.115pt}}
\multiput(315.17,363.68)(13.000,-30.685){2}{\rule{0.400pt}{0.558pt}}
\multiput(329.58,329.13)(0.492,-1.056){21}{\rule{0.119pt}{0.933pt}}
\multiput(328.17,331.06)(12.000,-23.063){2}{\rule{0.400pt}{0.467pt}}
\multiput(341.58,305.54)(0.493,-0.616){23}{\rule{0.119pt}{0.592pt}}
\multiput(340.17,306.77)(13.000,-14.771){2}{\rule{0.400pt}{0.296pt}}
\multiput(354.00,290.93)(0.728,-0.489){15}{\rule{0.678pt}{0.118pt}}
\multiput(354.00,291.17)(11.593,-9.000){2}{\rule{0.339pt}{0.400pt}}
\put(367,281.67){\rule{3.132pt}{0.400pt}}
\multiput(367.00,282.17)(6.500,-1.000){2}{\rule{1.566pt}{0.400pt}}
\multiput(380.00,282.59)(1.267,0.477){7}{\rule{1.060pt}{0.115pt}}
\multiput(380.00,281.17)(9.800,5.000){2}{\rule{0.530pt}{0.400pt}}
\multiput(392.00,287.58)(0.539,0.492){21}{\rule{0.533pt}{0.119pt}}
\multiput(392.00,286.17)(11.893,12.000){2}{\rule{0.267pt}{0.400pt}}
\multiput(405.58,299.00)(0.493,0.655){23}{\rule{0.119pt}{0.623pt}}
\multiput(404.17,299.00)(13.000,15.707){2}{\rule{0.400pt}{0.312pt}}
\multiput(418.58,316.00)(0.493,0.814){23}{\rule{0.119pt}{0.746pt}}
\multiput(417.17,316.00)(13.000,19.451){2}{\rule{0.400pt}{0.373pt}}
\multiput(431.58,337.00)(0.492,1.099){21}{\rule{0.119pt}{0.967pt}}
\multiput(430.17,337.00)(12.000,23.994){2}{\rule{0.400pt}{0.483pt}}
\multiput(443.58,363.00)(0.493,1.091){23}{\rule{0.119pt}{0.962pt}}
\multiput(442.17,363.00)(13.000,26.004){2}{\rule{0.400pt}{0.481pt}}
\multiput(456.58,391.00)(0.493,1.171){23}{\rule{0.119pt}{1.023pt}}
\multiput(455.17,391.00)(13.000,27.877){2}{\rule{0.400pt}{0.512pt}}
\multiput(469.58,421.00)(0.492,1.272){21}{\rule{0.119pt}{1.100pt}}
\multiput(468.17,421.00)(12.000,27.717){2}{\rule{0.400pt}{0.550pt}}
\multiput(481.58,451.00)(0.493,1.091){23}{\rule{0.119pt}{0.962pt}}
\multiput(480.17,451.00)(13.000,26.004){2}{\rule{0.400pt}{0.481pt}}
\multiput(494.58,479.00)(0.493,0.972){23}{\rule{0.119pt}{0.869pt}}
\multiput(493.17,479.00)(13.000,23.196){2}{\rule{0.400pt}{0.435pt}}
\multiput(507.58,504.00)(0.493,0.853){23}{\rule{0.119pt}{0.777pt}}
\multiput(506.17,504.00)(13.000,20.387){2}{\rule{0.400pt}{0.388pt}}
\multiput(520.58,526.00)(0.492,0.669){21}{\rule{0.119pt}{0.633pt}}
\multiput(519.17,526.00)(12.000,14.685){2}{\rule{0.400pt}{0.317pt}}
\multiput(532.00,542.58)(0.497,0.493){23}{\rule{0.500pt}{0.119pt}}
\multiput(532.00,541.17)(11.962,13.000){2}{\rule{0.250pt}{0.400pt}}
\multiput(545.00,555.59)(0.950,0.485){11}{\rule{0.843pt}{0.117pt}}
\multiput(545.00,554.17)(11.251,7.000){2}{\rule{0.421pt}{0.400pt}}
\multiput(558.00,562.61)(2.695,0.447){3}{\rule{1.833pt}{0.108pt}}
\multiput(558.00,561.17)(9.195,3.000){2}{\rule{0.917pt}{0.400pt}}
\multiput(583.00,563.94)(1.797,-0.468){5}{\rule{1.400pt}{0.113pt}}
\multiput(583.00,564.17)(10.094,-4.000){2}{\rule{0.700pt}{0.400pt}}
\multiput(596.00,559.93)(1.123,-0.482){9}{\rule{0.967pt}{0.116pt}}
\multiput(596.00,560.17)(10.994,-6.000){2}{\rule{0.483pt}{0.400pt}}
\multiput(609.00,553.93)(0.758,-0.488){13}{\rule{0.700pt}{0.117pt}}
\multiput(609.00,554.17)(10.547,-8.000){2}{\rule{0.350pt}{0.400pt}}
\multiput(621.00,545.93)(0.824,-0.488){13}{\rule{0.750pt}{0.117pt}}
\multiput(621.00,546.17)(11.443,-8.000){2}{\rule{0.375pt}{0.400pt}}
\multiput(634.00,537.92)(0.652,-0.491){17}{\rule{0.620pt}{0.118pt}}
\multiput(634.00,538.17)(11.713,-10.000){2}{\rule{0.310pt}{0.400pt}}
\multiput(647.00,527.92)(0.652,-0.491){17}{\rule{0.620pt}{0.118pt}}
\multiput(647.00,528.17)(11.713,-10.000){2}{\rule{0.310pt}{0.400pt}}
\multiput(660.00,517.93)(0.669,-0.489){15}{\rule{0.633pt}{0.118pt}}
\multiput(660.00,518.17)(10.685,-9.000){2}{\rule{0.317pt}{0.400pt}}
\multiput(672.00,508.93)(0.824,-0.488){13}{\rule{0.750pt}{0.117pt}}
\multiput(672.00,509.17)(11.443,-8.000){2}{\rule{0.375pt}{0.400pt}}
\multiput(685.00,500.93)(0.824,-0.488){13}{\rule{0.750pt}{0.117pt}}
\multiput(685.00,501.17)(11.443,-8.000){2}{\rule{0.375pt}{0.400pt}}
\multiput(698.00,492.93)(1.123,-0.482){9}{\rule{0.967pt}{0.116pt}}
\multiput(698.00,493.17)(10.994,-6.000){2}{\rule{0.483pt}{0.400pt}}
\multiput(711.00,486.93)(1.033,-0.482){9}{\rule{0.900pt}{0.116pt}}
\multiput(711.00,487.17)(10.132,-6.000){2}{\rule{0.450pt}{0.400pt}}
\multiput(723.00,480.94)(1.797,-0.468){5}{\rule{1.400pt}{0.113pt}}
\multiput(723.00,481.17)(10.094,-4.000){2}{\rule{0.700pt}{0.400pt}}
\multiput(736.00,476.95)(2.695,-0.447){3}{\rule{1.833pt}{0.108pt}}
\multiput(736.00,477.17)(9.195,-3.000){2}{\rule{0.917pt}{0.400pt}}
\put(749,473.67){\rule{2.891pt}{0.400pt}}
\multiput(749.00,474.17)(6.000,-1.000){2}{\rule{1.445pt}{0.400pt}}
\put(761,472.67){\rule{3.132pt}{0.400pt}}
\multiput(761.00,473.17)(6.500,-1.000){2}{\rule{1.566pt}{0.400pt}}
\put(774,472.67){\rule{3.132pt}{0.400pt}}
\multiput(774.00,472.17)(6.500,1.000){2}{\rule{1.566pt}{0.400pt}}
\put(787,473.67){\rule{3.132pt}{0.400pt}}
\multiput(787.00,473.17)(6.500,1.000){2}{\rule{1.566pt}{0.400pt}}
\put(800,475.17){\rule{2.500pt}{0.400pt}}
\multiput(800.00,474.17)(6.811,2.000){2}{\rule{1.250pt}{0.400pt}}
\put(812,477.17){\rule{2.700pt}{0.400pt}}
\multiput(812.00,476.17)(7.396,2.000){2}{\rule{1.350pt}{0.400pt}}
\multiput(825.00,479.61)(2.695,0.447){3}{\rule{1.833pt}{0.108pt}}
\multiput(825.00,478.17)(9.195,3.000){2}{\rule{0.917pt}{0.400pt}}
\multiput(838.00,482.61)(2.695,0.447){3}{\rule{1.833pt}{0.108pt}}
\multiput(838.00,481.17)(9.195,3.000){2}{\rule{0.917pt}{0.400pt}}
\multiput(851.00,485.61)(2.472,0.447){3}{\rule{1.700pt}{0.108pt}}
\multiput(851.00,484.17)(8.472,3.000){2}{\rule{0.850pt}{0.400pt}}
\multiput(863.00,488.61)(2.695,0.447){3}{\rule{1.833pt}{0.108pt}}
\multiput(863.00,487.17)(9.195,3.000){2}{\rule{0.917pt}{0.400pt}}
\put(876,491.17){\rule{2.700pt}{0.400pt}}
\multiput(876.00,490.17)(7.396,2.000){2}{\rule{1.350pt}{0.400pt}}
\multiput(889.00,493.61)(2.472,0.447){3}{\rule{1.700pt}{0.108pt}}
\multiput(889.00,492.17)(8.472,3.000){2}{\rule{0.850pt}{0.400pt}}
\put(901,496.17){\rule{2.700pt}{0.400pt}}
\multiput(901.00,495.17)(7.396,2.000){2}{\rule{1.350pt}{0.400pt}}
\put(914,497.67){\rule{3.132pt}{0.400pt}}
\multiput(914.00,497.17)(6.500,1.000){2}{\rule{1.566pt}{0.400pt}}
\put(927,498.67){\rule{3.132pt}{0.400pt}}
\multiput(927.00,498.17)(6.500,1.000){2}{\rule{1.566pt}{0.400pt}}
\put(940,499.67){\rule{2.891pt}{0.400pt}}
\multiput(940.00,499.17)(6.000,1.000){2}{\rule{1.445pt}{0.400pt}}
\put(952,500.67){\rule{3.132pt}{0.400pt}}
\multiput(952.00,500.17)(6.500,1.000){2}{\rule{1.566pt}{0.400pt}}
\put(571.0,565.0){\rule[-0.200pt]{2.891pt}{0.400pt}}
\put(978,500.67){\rule{3.132pt}{0.400pt}}
\multiput(978.00,501.17)(6.500,-1.000){2}{\rule{1.566pt}{0.400pt}}
\put(965.0,502.0){\rule[-0.200pt]{3.132pt}{0.400pt}}
\put(1003,499.67){\rule{3.132pt}{0.400pt}}
\multiput(1003.00,500.17)(6.500,-1.000){2}{\rule{1.566pt}{0.400pt}}
\put(1016,498.67){\rule{3.132pt}{0.400pt}}
\multiput(1016.00,499.17)(6.500,-1.000){2}{\rule{1.566pt}{0.400pt}}
\put(991.0,501.0){\rule[-0.200pt]{2.891pt}{0.400pt}}
\put(1041,497.67){\rule{3.132pt}{0.400pt}}
\multiput(1041.00,498.17)(6.500,-1.000){2}{\rule{1.566pt}{0.400pt}}
\put(1054,496.67){\rule{3.132pt}{0.400pt}}
\multiput(1054.00,497.17)(6.500,-1.000){2}{\rule{1.566pt}{0.400pt}}
\put(1067,495.67){\rule{3.132pt}{0.400pt}}
\multiput(1067.00,496.17)(6.500,-1.000){2}{\rule{1.566pt}{0.400pt}}
\put(1080,494.67){\rule{2.891pt}{0.400pt}}
\multiput(1080.00,495.17)(6.000,-1.000){2}{\rule{1.445pt}{0.400pt}}
\put(1029.0,499.0){\rule[-0.200pt]{2.891pt}{0.400pt}}
\put(1105,493.67){\rule{3.132pt}{0.400pt}}
\multiput(1105.00,494.17)(6.500,-1.000){2}{\rule{1.566pt}{0.400pt}}
\put(1092.0,495.0){\rule[-0.200pt]{3.132pt}{0.400pt}}
\put(1131,492.67){\rule{2.891pt}{0.400pt}}
\multiput(1131.00,493.17)(6.000,-1.000){2}{\rule{1.445pt}{0.400pt}}
\put(1118.0,494.0){\rule[-0.200pt]{3.132pt}{0.400pt}}
\put(1220,492.67){\rule{2.891pt}{0.400pt}}
\multiput(1220.00,492.17)(6.000,1.000){2}{\rule{1.445pt}{0.400pt}}
\put(1143.0,493.0){\rule[-0.200pt]{18.549pt}{0.400pt}}
\put(1258,493.67){\rule{3.132pt}{0.400pt}}
\multiput(1258.00,493.17)(6.500,1.000){2}{\rule{1.566pt}{0.400pt}}
\put(1232.0,494.0){\rule[-0.200pt]{6.263pt}{0.400pt}}
\put(1334,494.67){\rule{3.132pt}{0.400pt}}
\multiput(1334.00,494.17)(6.500,1.000){2}{\rule{1.566pt}{0.400pt}}
\put(1271.0,495.0){\rule[-0.200pt]{15.177pt}{0.400pt}}
\put(1411,494.67){\rule{2.891pt}{0.400pt}}
\multiput(1411.00,495.17)(6.000,-1.000){2}{\rule{1.445pt}{0.400pt}}
\put(1347.0,496.0){\rule[-0.200pt]{15.418pt}{0.400pt}}
\put(1423.0,495.0){\rule[-0.200pt]{3.132pt}{0.400pt}}
\sbox{\plotpoint}{\rule[-0.500pt]{1.000pt}{1.000pt}}%
\put(1306,722){\makebox(0,0)[r]{$P_{001}(t)$}}
\multiput(1328,722)(20.756,0.000){4}{\usebox{\plotpoint}}
\put(1394,722){\usebox{\plotpoint}}
\put(176,113){\usebox{\plotpoint}}
\put(176.00,113.00){\usebox{\plotpoint}}
\put(196.73,113.64){\usebox{\plotpoint}}
\multiput(201,114)(19.372,7.451){0}{\usebox{\plotpoint}}
\put(215.94,120.64){\usebox{\plotpoint}}
\put(230.30,135.33){\usebox{\plotpoint}}
\multiput(240,151)(7.093,19.506){2}{\usebox{\plotpoint}}
\multiput(252,184)(5.322,20.061){3}{\usebox{\plotpoint}}
\multiput(265,233)(4.195,20.327){3}{\usebox{\plotpoint}}
\multiput(278,296)(3.545,20.451){3}{\usebox{\plotpoint}}
\multiput(291,371)(3.005,20.537){4}{\usebox{\plotpoint}}
\multiput(303,453)(3.370,20.480){4}{\usebox{\plotpoint}}
\multiput(316,532)(3.897,20.386){3}{\usebox{\plotpoint}}
\multiput(329,600)(4.583,20.243){3}{\usebox{\plotpoint}}
\multiput(341,653)(7.227,19.457){2}{\usebox{\plotpoint}}
\put(361.67,698.03){\usebox{\plotpoint}}
\put(378.67,707.69){\usebox{\plotpoint}}
\multiput(380,708)(15.945,-13.287){0}{\usebox{\plotpoint}}
\put(394.29,695.00){\usebox{\plotpoint}}
\multiput(405,681)(9.885,-18.250){2}{\usebox{\plotpoint}}
\put(425.52,641.38){\usebox{\plotpoint}}
\multiput(431,630)(7.936,-19.178){2}{\usebox{\plotpoint}}
\put(450.47,584.34){\usebox{\plotpoint}}
\multiput(456,572)(8.740,-18.825){2}{\usebox{\plotpoint}}
\put(476.26,527.67){\usebox{\plotpoint}}
\put(485.47,509.09){\usebox{\plotpoint}}
\multiput(494,494)(10.925,-17.648){2}{\usebox{\plotpoint}}
\put(519.29,456.93){\usebox{\plotpoint}}
\multiput(520,456)(14.078,-15.251){0}{\usebox{\plotpoint}}
\put(533.50,441.85){\usebox{\plotpoint}}
\put(550.67,430.38){\usebox{\plotpoint}}
\put(570.54,425.07){\usebox{\plotpoint}}
\multiput(571,425)(20.756,0.000){0}{\usebox{\plotpoint}}
\put(591.07,426.86){\usebox{\plotpoint}}
\multiput(596,428)(19.372,7.451){0}{\usebox{\plotpoint}}
\put(610.47,433.98){\usebox{\plotpoint}}
\put(627.90,445.25){\usebox{\plotpoint}}
\put(645.18,456.74){\usebox{\plotpoint}}
\multiput(647,458)(16.451,12.655){0}{\usebox{\plotpoint}}
\put(661.71,469.28){\usebox{\plotpoint}}
\put(678.49,481.49){\usebox{\plotpoint}}
\put(695.93,492.73){\usebox{\plotpoint}}
\multiput(698,494)(18.275,9.840){0}{\usebox{\plotpoint}}
\put(714.19,502.59){\usebox{\plotpoint}}
\put(733.42,510.21){\usebox{\plotpoint}}
\multiput(736,511)(20.224,4.667){0}{\usebox{\plotpoint}}
\put(753.65,514.78){\usebox{\plotpoint}}
\multiput(761,516)(20.694,1.592){0}{\usebox{\plotpoint}}
\put(774.27,516.98){\usebox{\plotpoint}}
\put(794.96,515.39){\usebox{\plotpoint}}
\multiput(800,515)(20.473,-3.412){0}{\usebox{\plotpoint}}
\put(815.45,512.20){\usebox{\plotpoint}}
\put(835.82,508.33){\usebox{\plotpoint}}
\multiput(838,508)(20.224,-4.667){0}{\usebox{\plotpoint}}
\put(856.06,503.74){\usebox{\plotpoint}}
\multiput(863,502)(20.224,-4.667){0}{\usebox{\plotpoint}}
\put(876.25,498.94){\usebox{\plotpoint}}
\put(896.57,494.74){\usebox{\plotpoint}}
\multiput(901,494)(20.514,-3.156){0}{\usebox{\plotpoint}}
\put(917.10,491.76){\usebox{\plotpoint}}
\put(937.79,490.17){\usebox{\plotpoint}}
\multiput(940,490)(20.684,-1.724){0}{\usebox{\plotpoint}}
\put(958.48,488.50){\usebox{\plotpoint}}
\multiput(965,488)(20.756,0.000){0}{\usebox{\plotpoint}}
\put(979.21,488.09){\usebox{\plotpoint}}
\put(999.93,489.00){\usebox{\plotpoint}}
\multiput(1003,489)(20.694,1.592){0}{\usebox{\plotpoint}}
\put(1020.64,490.36){\usebox{\plotpoint}}
\multiput(1029,491)(20.756,0.000){0}{\usebox{\plotpoint}}
\put(1041.37,491.03){\usebox{\plotpoint}}
\put(1062.06,492.62){\usebox{\plotpoint}}
\multiput(1067,493)(20.694,1.592){0}{\usebox{\plotpoint}}
\put(1082.75,494.23){\usebox{\plotpoint}}
\put(1103.48,495.00){\usebox{\plotpoint}}
\multiput(1105,495)(20.694,1.592){0}{\usebox{\plotpoint}}
\put(1124.20,496.00){\usebox{\plotpoint}}
\multiput(1131,496)(20.684,1.724){0}{\usebox{\plotpoint}}
\put(1144.91,497.00){\usebox{\plotpoint}}
\put(1165.66,497.00){\usebox{\plotpoint}}
\multiput(1169,497)(20.756,0.000){0}{\usebox{\plotpoint}}
\put(1186.42,497.00){\usebox{\plotpoint}}
\multiput(1194,497)(20.756,0.000){0}{\usebox{\plotpoint}}
\put(1207.18,497.00){\usebox{\plotpoint}}
\put(1227.90,496.34){\usebox{\plotpoint}}
\multiput(1232,496)(20.756,0.000){0}{\usebox{\plotpoint}}
\put(1248.65,496.00){\usebox{\plotpoint}}
\put(1269.37,495.13){\usebox{\plotpoint}}
\multiput(1271,495)(20.756,0.000){0}{\usebox{\plotpoint}}
\put(1290.12,495.00){\usebox{\plotpoint}}
\multiput(1296,495)(20.756,0.000){0}{\usebox{\plotpoint}}
\put(1310.87,495.00){\usebox{\plotpoint}}
\put(1331.63,495.00){\usebox{\plotpoint}}
\multiput(1334,495)(20.694,-1.592){0}{\usebox{\plotpoint}}
\put(1352.35,494.00){\usebox{\plotpoint}}
\multiput(1360,494)(20.756,0.000){0}{\usebox{\plotpoint}}
\put(1373.10,494.00){\usebox{\plotpoint}}
\put(1393.86,494.00){\usebox{\plotpoint}}
\multiput(1398,494)(20.756,0.000){0}{\usebox{\plotpoint}}
\put(1414.60,494.30){\usebox{\plotpoint}}
\put(1435.33,495.00){\usebox{\plotpoint}}
\put(1436,495){\usebox{\plotpoint}}
\end{picture}

%% file: fig3.tex
\setlength{\unitlength}{0.240900pt}
\ifx\plotpoint\undefined\newsavebox{\plotpoint}\fi
\sbox{\plotpoint}{\rule[-0.200pt]{0.400pt}{0.400pt}}%
\begin{picture}(1500,900)(0,0)
\font\gnuplot=cmr10 at 10pt
\gnuplot
\sbox{\plotpoint}{\rule[-0.200pt]{0.400pt}{0.400pt}}%
\put(220.0,113.0){\rule[-0.200pt]{292.934pt}{0.400pt}}
\put(220.0,113.0){\rule[-0.200pt]{0.400pt}{184.048pt}}
\put(220.0,113.0){\rule[-0.200pt]{4.818pt}{0.400pt}}
\put(198,113){\makebox(0,0)[r]{0}}
\put(1416.0,113.0){\rule[-0.200pt]{4.818pt}{0.400pt}}
\put(220.0,189.0){\rule[-0.200pt]{4.818pt}{0.400pt}}
\put(198,189){\makebox(0,0)[r]{0.05}}
\put(1416.0,189.0){\rule[-0.200pt]{4.818pt}{0.400pt}}
\put(220.0,266.0){\rule[-0.200pt]{4.818pt}{0.400pt}}
\put(198,266){\makebox(0,0)[r]{0.1}}
\put(1416.0,266.0){\rule[-0.200pt]{4.818pt}{0.400pt}}
\put(220.0,342.0){\rule[-0.200pt]{4.818pt}{0.400pt}}
\put(198,342){\makebox(0,0)[r]{0.15}}
\put(1416.0,342.0){\rule[-0.200pt]{4.818pt}{0.400pt}}
\put(220.0,419.0){\rule[-0.200pt]{4.818pt}{0.400pt}}
\put(198,419){\makebox(0,0)[r]{0.2}}
\put(1416.0,419.0){\rule[-0.200pt]{4.818pt}{0.400pt}}
\put(220.0,495.0){\rule[-0.200pt]{4.818pt}{0.400pt}}
\put(198,495){\makebox(0,0)[r]{0.25}}
\put(1416.0,495.0){\rule[-0.200pt]{4.818pt}{0.400pt}}
\put(220.0,571.0){\rule[-0.200pt]{4.818pt}{0.400pt}}
\put(198,571){\makebox(0,0)[r]{0.3}}
\put(1416.0,571.0){\rule[-0.200pt]{4.818pt}{0.400pt}}
\put(220.0,648.0){\rule[-0.200pt]{4.818pt}{0.400pt}}
\put(198,648){\makebox(0,0)[r]{0.35}}
\put(1416.0,648.0){\rule[-0.200pt]{4.818pt}{0.400pt}}
\put(220.0,724.0){\rule[-0.200pt]{4.818pt}{0.400pt}}
\put(198,724){\makebox(0,0)[r]{0.4}}
\put(1416.0,724.0){\rule[-0.200pt]{4.818pt}{0.400pt}}
\put(220.0,801.0){\rule[-0.200pt]{4.818pt}{0.400pt}}
\put(198,801){\makebox(0,0)[r]{0.45}}
\put(1416.0,801.0){\rule[-0.200pt]{4.818pt}{0.400pt}}
\put(220.0,877.0){\rule[-0.200pt]{4.818pt}{0.400pt}}
\put(198,877){\makebox(0,0)[r]{0.5}}
\put(1416.0,877.0){\rule[-0.200pt]{4.818pt}{0.400pt}}
\put(220.0,113.0){\rule[-0.200pt]{0.400pt}{4.818pt}}
\put(220,68){\makebox(0,0){0}}
\put(220.0,857.0){\rule[-0.200pt]{0.400pt}{4.818pt}}
\put(463.0,113.0){\rule[-0.200pt]{0.400pt}{4.818pt}}
\put(463,68){\makebox(0,0){1}}
\put(463.0,857.0){\rule[-0.200pt]{0.400pt}{4.818pt}}
\put(706.0,113.0){\rule[-0.200pt]{0.400pt}{4.818pt}}
\put(706,68){\makebox(0,0){2}}
\put(706.0,857.0){\rule[-0.200pt]{0.400pt}{4.818pt}}
\put(950.0,113.0){\rule[-0.200pt]{0.400pt}{4.818pt}}
\put(950,68){\makebox(0,0){3}}
\put(950.0,857.0){\rule[-0.200pt]{0.400pt}{4.818pt}}
\put(1193.0,113.0){\rule[-0.200pt]{0.400pt}{4.818pt}}
\put(1193,68){\makebox(0,0){4}}
\put(1193.0,857.0){\rule[-0.200pt]{0.400pt}{4.818pt}}
\put(1436.0,113.0){\rule[-0.200pt]{0.400pt}{4.818pt}}
\put(1436,68){\makebox(0,0){5}}
\put(1436.0,857.0){\rule[-0.200pt]{0.400pt}{4.818pt}}
\put(220.0,113.0){\rule[-0.200pt]{292.934pt}{0.400pt}}
\put(1436.0,113.0){\rule[-0.200pt]{0.400pt}{184.048pt}}
\put(220.0,877.0){\rule[-0.200pt]{292.934pt}{0.400pt}}
\put(45,495){\makebox(0,0){$P_{cav}$}}
\put(828,23){\makebox(0,0){$t~[1/g]$}}
\put(220.0,113.0){\rule[-0.200pt]{0.400pt}{184.048pt}}
\put(220,113){\usebox{\plotpoint}}
\put(232,112.67){\rule{3.132pt}{0.400pt}}
\multiput(232.00,112.17)(6.500,1.000){2}{\rule{1.566pt}{0.400pt}}
\put(245,114.17){\rule{2.500pt}{0.400pt}}
\multiput(245.00,113.17)(6.811,2.000){2}{\rule{1.250pt}{0.400pt}}
\multiput(257.00,116.60)(1.651,0.468){5}{\rule{1.300pt}{0.113pt}}
\multiput(257.00,115.17)(9.302,4.000){2}{\rule{0.650pt}{0.400pt}}
\multiput(269.00,120.59)(1.033,0.482){9}{\rule{0.900pt}{0.116pt}}
\multiput(269.00,119.17)(10.132,6.000){2}{\rule{0.450pt}{0.400pt}}
\multiput(281.00,126.59)(0.728,0.489){15}{\rule{0.678pt}{0.118pt}}
\multiput(281.00,125.17)(11.593,9.000){2}{\rule{0.339pt}{0.400pt}}
\multiput(294.00,135.58)(0.543,0.492){19}{\rule{0.536pt}{0.118pt}}
\multiput(294.00,134.17)(10.887,11.000){2}{\rule{0.268pt}{0.400pt}}
\multiput(306.58,146.00)(0.492,0.582){21}{\rule{0.119pt}{0.567pt}}
\multiput(305.17,146.00)(12.000,12.824){2}{\rule{0.400pt}{0.283pt}}
\multiput(318.58,160.00)(0.493,0.655){23}{\rule{0.119pt}{0.623pt}}
\multiput(317.17,160.00)(13.000,15.707){2}{\rule{0.400pt}{0.312pt}}
\multiput(331.58,177.00)(0.492,0.798){21}{\rule{0.119pt}{0.733pt}}
\multiput(330.17,177.00)(12.000,17.478){2}{\rule{0.400pt}{0.367pt}}
\multiput(343.58,196.00)(0.492,0.884){21}{\rule{0.119pt}{0.800pt}}
\multiput(342.17,196.00)(12.000,19.340){2}{\rule{0.400pt}{0.400pt}}
\multiput(355.58,217.00)(0.492,1.013){21}{\rule{0.119pt}{0.900pt}}
\multiput(354.17,217.00)(12.000,22.132){2}{\rule{0.400pt}{0.450pt}}
\multiput(367.58,241.00)(0.493,1.012){23}{\rule{0.119pt}{0.900pt}}
\multiput(366.17,241.00)(13.000,24.132){2}{\rule{0.400pt}{0.450pt}}
\multiput(380.58,267.00)(0.492,1.142){21}{\rule{0.119pt}{1.000pt}}
\multiput(379.17,267.00)(12.000,24.924){2}{\rule{0.400pt}{0.500pt}}
\multiput(392.58,294.00)(0.492,1.229){21}{\rule{0.119pt}{1.067pt}}
\multiput(391.17,294.00)(12.000,26.786){2}{\rule{0.400pt}{0.533pt}}
\multiput(404.58,323.00)(0.493,1.171){23}{\rule{0.119pt}{1.023pt}}
\multiput(403.17,323.00)(13.000,27.877){2}{\rule{0.400pt}{0.512pt}}
\multiput(417.58,353.00)(0.492,1.272){21}{\rule{0.119pt}{1.100pt}}
\multiput(416.17,353.00)(12.000,27.717){2}{\rule{0.400pt}{0.550pt}}
\multiput(429.58,383.00)(0.492,1.315){21}{\rule{0.119pt}{1.133pt}}
\multiput(428.17,383.00)(12.000,28.648){2}{\rule{0.400pt}{0.567pt}}
\multiput(441.58,414.00)(0.492,1.315){21}{\rule{0.119pt}{1.133pt}}
\multiput(440.17,414.00)(12.000,28.648){2}{\rule{0.400pt}{0.567pt}}
\multiput(453.58,445.00)(0.493,1.210){23}{\rule{0.119pt}{1.054pt}}
\multiput(452.17,445.00)(13.000,28.813){2}{\rule{0.400pt}{0.527pt}}
\multiput(466.58,476.00)(0.492,1.272){21}{\rule{0.119pt}{1.100pt}}
\multiput(465.17,476.00)(12.000,27.717){2}{\rule{0.400pt}{0.550pt}}
\multiput(478.58,506.00)(0.492,1.229){21}{\rule{0.119pt}{1.067pt}}
\multiput(477.17,506.00)(12.000,26.786){2}{\rule{0.400pt}{0.533pt}}
\multiput(490.58,535.00)(0.493,1.091){23}{\rule{0.119pt}{0.962pt}}
\multiput(489.17,535.00)(13.000,26.004){2}{\rule{0.400pt}{0.481pt}}
\multiput(503.58,563.00)(0.492,1.142){21}{\rule{0.119pt}{1.000pt}}
\multiput(502.17,563.00)(12.000,24.924){2}{\rule{0.400pt}{0.500pt}}
\multiput(515.58,590.00)(0.492,1.056){21}{\rule{0.119pt}{0.933pt}}
\multiput(514.17,590.00)(12.000,23.063){2}{\rule{0.400pt}{0.467pt}}
\multiput(527.58,615.00)(0.492,1.013){21}{\rule{0.119pt}{0.900pt}}
\multiput(526.17,615.00)(12.000,22.132){2}{\rule{0.400pt}{0.450pt}}
\multiput(539.58,639.00)(0.493,0.853){23}{\rule{0.119pt}{0.777pt}}
\multiput(538.17,639.00)(13.000,20.387){2}{\rule{0.400pt}{0.388pt}}
\multiput(552.58,661.00)(0.492,0.884){21}{\rule{0.119pt}{0.800pt}}
\multiput(551.17,661.00)(12.000,19.340){2}{\rule{0.400pt}{0.400pt}}
\multiput(564.58,682.00)(0.492,0.755){21}{\rule{0.119pt}{0.700pt}}
\multiput(563.17,682.00)(12.000,16.547){2}{\rule{0.400pt}{0.350pt}}
\multiput(576.58,700.00)(0.492,0.712){21}{\rule{0.119pt}{0.667pt}}
\multiput(575.17,700.00)(12.000,15.616){2}{\rule{0.400pt}{0.333pt}}
\multiput(588.58,717.00)(0.493,0.576){23}{\rule{0.119pt}{0.562pt}}
\multiput(587.17,717.00)(13.000,13.834){2}{\rule{0.400pt}{0.281pt}}
\multiput(601.58,732.00)(0.492,0.582){21}{\rule{0.119pt}{0.567pt}}
\multiput(600.17,732.00)(12.000,12.824){2}{\rule{0.400pt}{0.283pt}}
\multiput(613.00,746.58)(0.543,0.492){19}{\rule{0.536pt}{0.118pt}}
\multiput(613.00,745.17)(10.887,11.000){2}{\rule{0.268pt}{0.400pt}}
\multiput(625.00,757.58)(0.652,0.491){17}{\rule{0.620pt}{0.118pt}}
\multiput(625.00,756.17)(11.713,10.000){2}{\rule{0.310pt}{0.400pt}}
\multiput(638.00,767.59)(0.669,0.489){15}{\rule{0.633pt}{0.118pt}}
\multiput(638.00,766.17)(10.685,9.000){2}{\rule{0.317pt}{0.400pt}}
\multiput(650.00,776.59)(0.874,0.485){11}{\rule{0.786pt}{0.117pt}}
\multiput(650.00,775.17)(10.369,7.000){2}{\rule{0.393pt}{0.400pt}}
\multiput(662.00,783.59)(1.033,0.482){9}{\rule{0.900pt}{0.116pt}}
\multiput(662.00,782.17)(10.132,6.000){2}{\rule{0.450pt}{0.400pt}}
\multiput(674.00,789.60)(1.797,0.468){5}{\rule{1.400pt}{0.113pt}}
\multiput(674.00,788.17)(10.094,4.000){2}{\rule{0.700pt}{0.400pt}}
\multiput(687.00,793.60)(1.651,0.468){5}{\rule{1.300pt}{0.113pt}}
\multiput(687.00,792.17)(9.302,4.000){2}{\rule{0.650pt}{0.400pt}}
\multiput(699.00,797.61)(2.472,0.447){3}{\rule{1.700pt}{0.108pt}}
\multiput(699.00,796.17)(8.472,3.000){2}{\rule{0.850pt}{0.400pt}}
\put(711,800.17){\rule{2.700pt}{0.400pt}}
\multiput(711.00,799.17)(7.396,2.000){2}{\rule{1.350pt}{0.400pt}}
\put(724,801.67){\rule{2.891pt}{0.400pt}}
\multiput(724.00,801.17)(6.000,1.000){2}{\rule{1.445pt}{0.400pt}}
\put(736,802.67){\rule{2.891pt}{0.400pt}}
\multiput(736.00,802.17)(6.000,1.000){2}{\rule{1.445pt}{0.400pt}}
\put(748,803.67){\rule{2.891pt}{0.400pt}}
\multiput(748.00,803.17)(6.000,1.000){2}{\rule{1.445pt}{0.400pt}}
\put(220.0,113.0){\rule[-0.200pt]{2.891pt}{0.400pt}}
\put(834,804.67){\rule{2.891pt}{0.400pt}}
\multiput(834.00,804.17)(6.000,1.000){2}{\rule{1.445pt}{0.400pt}}
\put(760.0,805.0){\rule[-0.200pt]{17.827pt}{0.400pt}}
\put(859,805.67){\rule{2.891pt}{0.400pt}}
\multiput(859.00,805.17)(6.000,1.000){2}{\rule{1.445pt}{0.400pt}}
\put(871,806.67){\rule{2.891pt}{0.400pt}}
\multiput(871.00,806.17)(6.000,1.000){2}{\rule{1.445pt}{0.400pt}}
\put(883,807.67){\rule{3.132pt}{0.400pt}}
\multiput(883.00,807.17)(6.500,1.000){2}{\rule{1.566pt}{0.400pt}}
\put(896,809.17){\rule{2.500pt}{0.400pt}}
\multiput(896.00,808.17)(6.811,2.000){2}{\rule{1.250pt}{0.400pt}}
\put(908,811.17){\rule{2.500pt}{0.400pt}}
\multiput(908.00,810.17)(6.811,2.000){2}{\rule{1.250pt}{0.400pt}}
\put(920,813.17){\rule{2.500pt}{0.400pt}}
\multiput(920.00,812.17)(6.811,2.000){2}{\rule{1.250pt}{0.400pt}}
\put(932,815.17){\rule{2.700pt}{0.400pt}}
\multiput(932.00,814.17)(7.396,2.000){2}{\rule{1.350pt}{0.400pt}}
\put(945,817.17){\rule{2.500pt}{0.400pt}}
\multiput(945.00,816.17)(6.811,2.000){2}{\rule{1.250pt}{0.400pt}}
\multiput(957.00,819.61)(2.472,0.447){3}{\rule{1.700pt}{0.108pt}}
\multiput(957.00,818.17)(8.472,3.000){2}{\rule{0.850pt}{0.400pt}}
\multiput(969.00,822.61)(2.695,0.447){3}{\rule{1.833pt}{0.108pt}}
\multiput(969.00,821.17)(9.195,3.000){2}{\rule{0.917pt}{0.400pt}}
\put(982,825.17){\rule{2.500pt}{0.400pt}}
\multiput(982.00,824.17)(6.811,2.000){2}{\rule{1.250pt}{0.400pt}}
\multiput(994.00,827.61)(2.472,0.447){3}{\rule{1.700pt}{0.108pt}}
\multiput(994.00,826.17)(8.472,3.000){2}{\rule{0.850pt}{0.400pt}}
\multiput(1006.00,830.61)(2.472,0.447){3}{\rule{1.700pt}{0.108pt}}
\multiput(1006.00,829.17)(8.472,3.000){2}{\rule{0.850pt}{0.400pt}}
\multiput(1018.00,833.61)(2.695,0.447){3}{\rule{1.833pt}{0.108pt}}
\multiput(1018.00,832.17)(9.195,3.000){2}{\rule{0.917pt}{0.400pt}}
\multiput(1031.00,836.61)(2.472,0.447){3}{\rule{1.700pt}{0.108pt}}
\multiput(1031.00,835.17)(8.472,3.000){2}{\rule{0.850pt}{0.400pt}}
\put(1043,839.17){\rule{2.500pt}{0.400pt}}
\multiput(1043.00,838.17)(6.811,2.000){2}{\rule{1.250pt}{0.400pt}}
\multiput(1055.00,841.61)(2.695,0.447){3}{\rule{1.833pt}{0.108pt}}
\multiput(1055.00,840.17)(9.195,3.000){2}{\rule{0.917pt}{0.400pt}}
\multiput(1068.00,844.61)(2.472,0.447){3}{\rule{1.700pt}{0.108pt}}
\multiput(1068.00,843.17)(8.472,3.000){2}{\rule{0.850pt}{0.400pt}}
\put(1080,847.17){\rule{2.500pt}{0.400pt}}
\multiput(1080.00,846.17)(6.811,2.000){2}{\rule{1.250pt}{0.400pt}}
\multiput(1092.00,849.61)(2.472,0.447){3}{\rule{1.700pt}{0.108pt}}
\multiput(1092.00,848.17)(8.472,3.000){2}{\rule{0.850pt}{0.400pt}}
\put(1104,852.17){\rule{2.700pt}{0.400pt}}
\multiput(1104.00,851.17)(7.396,2.000){2}{\rule{1.350pt}{0.400pt}}
\put(1117,854.17){\rule{2.500pt}{0.400pt}}
\multiput(1117.00,853.17)(6.811,2.000){2}{\rule{1.250pt}{0.400pt}}
\put(1129,856.17){\rule{2.500pt}{0.400pt}}
\multiput(1129.00,855.17)(6.811,2.000){2}{\rule{1.250pt}{0.400pt}}
\put(1141,858.17){\rule{2.500pt}{0.400pt}}
\multiput(1141.00,857.17)(6.811,2.000){2}{\rule{1.250pt}{0.400pt}}
\put(1153,859.67){\rule{3.132pt}{0.400pt}}
\multiput(1153.00,859.17)(6.500,1.000){2}{\rule{1.566pt}{0.400pt}}
\put(1166,861.17){\rule{2.500pt}{0.400pt}}
\multiput(1166.00,860.17)(6.811,2.000){2}{\rule{1.250pt}{0.400pt}}
\put(1178,862.67){\rule{2.891pt}{0.400pt}}
\multiput(1178.00,862.17)(6.000,1.000){2}{\rule{1.445pt}{0.400pt}}
\put(1190,863.67){\rule{3.132pt}{0.400pt}}
\multiput(1190.00,863.17)(6.500,1.000){2}{\rule{1.566pt}{0.400pt}}
\put(1203,864.67){\rule{2.891pt}{0.400pt}}
\multiput(1203.00,864.17)(6.000,1.000){2}{\rule{1.445pt}{0.400pt}}
\put(1215,865.67){\rule{2.891pt}{0.400pt}}
\multiput(1215.00,865.17)(6.000,1.000){2}{\rule{1.445pt}{0.400pt}}
\put(846.0,806.0){\rule[-0.200pt]{3.132pt}{0.400pt}}
\put(1239,866.67){\rule{3.132pt}{0.400pt}}
\multiput(1239.00,866.17)(6.500,1.000){2}{\rule{1.566pt}{0.400pt}}
\put(1227.0,867.0){\rule[-0.200pt]{2.891pt}{0.400pt}}
\put(1264,867.67){\rule{2.891pt}{0.400pt}}
\multiput(1264.00,867.17)(6.000,1.000){2}{\rule{1.445pt}{0.400pt}}
\put(1252.0,868.0){\rule[-0.200pt]{2.891pt}{0.400pt}}
\put(1276.0,869.0){\rule[-0.200pt]{38.544pt}{0.400pt}}
\end{picture}

%% file: fig4.tex
\setlength{\unitlength}{0.240900pt}
\ifx\plotpoint\undefined\newsavebox{\plotpoint}\fi
\sbox{\plotpoint}{\rule[-0.200pt]{0.400pt}{0.400pt}}%
\begin{picture}(1500,900)(0,0)
\font\gnuplot=cmr10 at 10pt
\gnuplot
\sbox{\plotpoint}{\rule[-0.200pt]{0.400pt}{0.400pt}}%
\put(220.0,113.0){\rule[-0.200pt]{4.818pt}{0.400pt}}
\put(198,113){\makebox(0,0)[r]{0.3}}
\put(1416.0,113.0){\rule[-0.200pt]{4.818pt}{0.400pt}}
\put(220.0,222.0){\rule[-0.200pt]{4.818pt}{0.400pt}}
\put(198,222){\makebox(0,0)[r]{0.4}}
\put(1416.0,222.0){\rule[-0.200pt]{4.818pt}{0.400pt}}
\put(220.0,331.0){\rule[-0.200pt]{4.818pt}{0.400pt}}
\put(198,331){\makebox(0,0)[r]{0.5}}
\put(1416.0,331.0){\rule[-0.200pt]{4.818pt}{0.400pt}}
\put(220.0,440.0){\rule[-0.200pt]{4.818pt}{0.400pt}}
\put(198,440){\makebox(0,0)[r]{0.6}}
\put(1416.0,440.0){\rule[-0.200pt]{4.818pt}{0.400pt}}
\put(220.0,550.0){\rule[-0.200pt]{4.818pt}{0.400pt}}
\put(198,550){\makebox(0,0)[r]{0.7}}
\put(1416.0,550.0){\rule[-0.200pt]{4.818pt}{0.400pt}}
\put(220.0,659.0){\rule[-0.200pt]{4.818pt}{0.400pt}}
\put(198,659){\makebox(0,0)[r]{0.8}}
\put(1416.0,659.0){\rule[-0.200pt]{4.818pt}{0.400pt}}
\put(220.0,768.0){\rule[-0.200pt]{4.818pt}{0.400pt}}
\put(198,768){\makebox(0,0)[r]{0.9}}
\put(1416.0,768.0){\rule[-0.200pt]{4.818pt}{0.400pt}}
\put(220.0,877.0){\rule[-0.200pt]{4.818pt}{0.400pt}}
\put(198,877){\makebox(0,0)[r]{1}}
\put(1416.0,877.0){\rule[-0.200pt]{4.818pt}{0.400pt}}
\put(331.0,113.0){\rule[-0.200pt]{0.400pt}{4.818pt}}
\put(331,68){\makebox(0,0){50}}
\put(331.0,857.0){\rule[-0.200pt]{0.400pt}{4.818pt}}
\put(453.0,113.0){\rule[-0.200pt]{0.400pt}{4.818pt}}
\put(453,68){\makebox(0,0){100}}
\put(453.0,857.0){\rule[-0.200pt]{0.400pt}{4.818pt}}
\put(576.0,113.0){\rule[-0.200pt]{0.400pt}{4.818pt}}
\put(576,68){\makebox(0,0){150}}
\put(576.0,857.0){\rule[-0.200pt]{0.400pt}{4.818pt}}
\put(699.0,113.0){\rule[-0.200pt]{0.400pt}{4.818pt}}
\put(699,68){\makebox(0,0){200}}
\put(699.0,857.0){\rule[-0.200pt]{0.400pt}{4.818pt}}
\put(822.0,113.0){\rule[-0.200pt]{0.400pt}{4.818pt}}
\put(822,68){\makebox(0,0){250}}
\put(822.0,857.0){\rule[-0.200pt]{0.400pt}{4.818pt}}
\put(945.0,113.0){\rule[-0.200pt]{0.400pt}{4.818pt}}
\put(945,68){\makebox(0,0){300}}
\put(945.0,857.0){\rule[-0.200pt]{0.400pt}{4.818pt}}
\put(1068.0,113.0){\rule[-0.200pt]{0.400pt}{4.818pt}}
\put(1068,68){\makebox(0,0){350}}
\put(1068.0,857.0){\rule[-0.200pt]{0.400pt}{4.818pt}}
\put(1190.0,113.0){\rule[-0.200pt]{0.400pt}{4.818pt}}
\put(1190,68){\makebox(0,0){400}}
\put(1190.0,857.0){\rule[-0.200pt]{0.400pt}{4.818pt}}
\put(1313.0,113.0){\rule[-0.200pt]{0.400pt}{4.818pt}}
\put(1313,68){\makebox(0,0){450}}
\put(1313.0,857.0){\rule[-0.200pt]{0.400pt}{4.818pt}}
\put(1436.0,113.0){\rule[-0.200pt]{0.400pt}{4.818pt}}
\put(1436,68){\makebox(0,0){500}}
\put(1436.0,857.0){\rule[-0.200pt]{0.400pt}{4.818pt}}
\put(220.0,113.0){\rule[-0.200pt]{292.934pt}{0.400pt}}
\put(1436.0,113.0){\rule[-0.200pt]{0.400pt}{184.048pt}}
\put(220.0,877.0){\rule[-0.200pt]{292.934pt}{0.400pt}}
\put(45,495){\makebox(0,0){$F_{asymp}$}}
\put(828,23){\makebox(0,0){$t~[1/g]$}}
\put(220.0,113.0){\rule[-0.200pt]{0.400pt}{184.048pt}}
\put(1306,812){\makebox(0,0)[r]{$\eta=1$}}
\put(1328.0,812.0){\rule[-0.200pt]{15.899pt}{0.400pt}}
\put(220,865){\usebox{\plotpoint}}
\multiput(220.00,863.92)(0.543,-0.492){19}{\rule{0.536pt}{0.118pt}}
\multiput(220.00,864.17)(10.887,-11.000){2}{\rule{0.268pt}{0.400pt}}
\multiput(232.00,852.92)(0.590,-0.492){19}{\rule{0.573pt}{0.118pt}}
\multiput(232.00,853.17)(11.811,-11.000){2}{\rule{0.286pt}{0.400pt}}
\multiput(245.00,841.92)(0.600,-0.491){17}{\rule{0.580pt}{0.118pt}}
\multiput(245.00,842.17)(10.796,-10.000){2}{\rule{0.290pt}{0.400pt}}
\multiput(257.00,831.92)(0.543,-0.492){19}{\rule{0.536pt}{0.118pt}}
\multiput(257.00,832.17)(10.887,-11.000){2}{\rule{0.268pt}{0.400pt}}
\multiput(269.00,820.92)(0.600,-0.491){17}{\rule{0.580pt}{0.118pt}}
\multiput(269.00,821.17)(10.796,-10.000){2}{\rule{0.290pt}{0.400pt}}
\multiput(281.00,810.92)(0.652,-0.491){17}{\rule{0.620pt}{0.118pt}}
\multiput(281.00,811.17)(11.713,-10.000){2}{\rule{0.310pt}{0.400pt}}
\multiput(294.00,800.92)(0.600,-0.491){17}{\rule{0.580pt}{0.118pt}}
\multiput(294.00,801.17)(10.796,-10.000){2}{\rule{0.290pt}{0.400pt}}
\multiput(306.00,790.92)(0.600,-0.491){17}{\rule{0.580pt}{0.118pt}}
\multiput(306.00,791.17)(10.796,-10.000){2}{\rule{0.290pt}{0.400pt}}
\multiput(318.00,780.92)(0.652,-0.491){17}{\rule{0.620pt}{0.118pt}}
\multiput(318.00,781.17)(11.713,-10.000){2}{\rule{0.310pt}{0.400pt}}
\multiput(331.00,770.92)(0.600,-0.491){17}{\rule{0.580pt}{0.118pt}}
\multiput(331.00,771.17)(10.796,-10.000){2}{\rule{0.290pt}{0.400pt}}
\multiput(343.00,760.92)(0.600,-0.491){17}{\rule{0.580pt}{0.118pt}}
\multiput(343.00,761.17)(10.796,-10.000){2}{\rule{0.290pt}{0.400pt}}
\multiput(355.00,750.93)(0.669,-0.489){15}{\rule{0.633pt}{0.118pt}}
\multiput(355.00,751.17)(10.685,-9.000){2}{\rule{0.317pt}{0.400pt}}
\multiput(367.00,741.92)(0.652,-0.491){17}{\rule{0.620pt}{0.118pt}}
\multiput(367.00,742.17)(11.713,-10.000){2}{\rule{0.310pt}{0.400pt}}
\multiput(380.00,731.93)(0.669,-0.489){15}{\rule{0.633pt}{0.118pt}}
\multiput(380.00,732.17)(10.685,-9.000){2}{\rule{0.317pt}{0.400pt}}
\multiput(392.00,722.92)(0.600,-0.491){17}{\rule{0.580pt}{0.118pt}}
\multiput(392.00,723.17)(10.796,-10.000){2}{\rule{0.290pt}{0.400pt}}
\multiput(404.00,712.93)(0.728,-0.489){15}{\rule{0.678pt}{0.118pt}}
\multiput(404.00,713.17)(11.593,-9.000){2}{\rule{0.339pt}{0.400pt}}
\multiput(417.00,703.93)(0.669,-0.489){15}{\rule{0.633pt}{0.118pt}}
\multiput(417.00,704.17)(10.685,-9.000){2}{\rule{0.317pt}{0.400pt}}
\multiput(429.00,694.93)(0.669,-0.489){15}{\rule{0.633pt}{0.118pt}}
\multiput(429.00,695.17)(10.685,-9.000){2}{\rule{0.317pt}{0.400pt}}
\multiput(441.00,685.93)(0.669,-0.489){15}{\rule{0.633pt}{0.118pt}}
\multiput(441.00,686.17)(10.685,-9.000){2}{\rule{0.317pt}{0.400pt}}
\multiput(453.00,676.93)(0.728,-0.489){15}{\rule{0.678pt}{0.118pt}}
\multiput(453.00,677.17)(11.593,-9.000){2}{\rule{0.339pt}{0.400pt}}
\multiput(466.00,667.93)(0.669,-0.489){15}{\rule{0.633pt}{0.118pt}}
\multiput(466.00,668.17)(10.685,-9.000){2}{\rule{0.317pt}{0.400pt}}
\multiput(478.00,658.93)(0.669,-0.489){15}{\rule{0.633pt}{0.118pt}}
\multiput(478.00,659.17)(10.685,-9.000){2}{\rule{0.317pt}{0.400pt}}
\multiput(490.00,649.93)(0.824,-0.488){13}{\rule{0.750pt}{0.117pt}}
\multiput(490.00,650.17)(11.443,-8.000){2}{\rule{0.375pt}{0.400pt}}
\multiput(503.00,641.93)(0.669,-0.489){15}{\rule{0.633pt}{0.118pt}}
\multiput(503.00,642.17)(10.685,-9.000){2}{\rule{0.317pt}{0.400pt}}
\multiput(515.00,632.93)(0.758,-0.488){13}{\rule{0.700pt}{0.117pt}}
\multiput(515.00,633.17)(10.547,-8.000){2}{\rule{0.350pt}{0.400pt}}
\multiput(527.00,624.93)(0.669,-0.489){15}{\rule{0.633pt}{0.118pt}}
\multiput(527.00,625.17)(10.685,-9.000){2}{\rule{0.317pt}{0.400pt}}
\multiput(539.00,615.93)(0.824,-0.488){13}{\rule{0.750pt}{0.117pt}}
\multiput(539.00,616.17)(11.443,-8.000){2}{\rule{0.375pt}{0.400pt}}
\multiput(552.00,607.93)(0.758,-0.488){13}{\rule{0.700pt}{0.117pt}}
\multiput(552.00,608.17)(10.547,-8.000){2}{\rule{0.350pt}{0.400pt}}
\multiput(564.00,599.93)(0.758,-0.488){13}{\rule{0.700pt}{0.117pt}}
\multiput(564.00,600.17)(10.547,-8.000){2}{\rule{0.350pt}{0.400pt}}
\multiput(576.00,591.93)(0.758,-0.488){13}{\rule{0.700pt}{0.117pt}}
\multiput(576.00,592.17)(10.547,-8.000){2}{\rule{0.350pt}{0.400pt}}
\multiput(588.00,583.93)(0.824,-0.488){13}{\rule{0.750pt}{0.117pt}}
\multiput(588.00,584.17)(11.443,-8.000){2}{\rule{0.375pt}{0.400pt}}
\multiput(601.00,575.93)(0.758,-0.488){13}{\rule{0.700pt}{0.117pt}}
\multiput(601.00,576.17)(10.547,-8.000){2}{\rule{0.350pt}{0.400pt}}
\multiput(613.00,567.93)(0.758,-0.488){13}{\rule{0.700pt}{0.117pt}}
\multiput(613.00,568.17)(10.547,-8.000){2}{\rule{0.350pt}{0.400pt}}
\multiput(625.00,559.93)(0.950,-0.485){11}{\rule{0.843pt}{0.117pt}}
\multiput(625.00,560.17)(11.251,-7.000){2}{\rule{0.421pt}{0.400pt}}
\multiput(638.00,552.93)(0.758,-0.488){13}{\rule{0.700pt}{0.117pt}}
\multiput(638.00,553.17)(10.547,-8.000){2}{\rule{0.350pt}{0.400pt}}
\multiput(650.00,544.93)(0.758,-0.488){13}{\rule{0.700pt}{0.117pt}}
\multiput(650.00,545.17)(10.547,-8.000){2}{\rule{0.350pt}{0.400pt}}
\multiput(662.00,536.93)(0.874,-0.485){11}{\rule{0.786pt}{0.117pt}}
\multiput(662.00,537.17)(10.369,-7.000){2}{\rule{0.393pt}{0.400pt}}
\multiput(674.00,529.93)(0.824,-0.488){13}{\rule{0.750pt}{0.117pt}}
\multiput(674.00,530.17)(11.443,-8.000){2}{\rule{0.375pt}{0.400pt}}
\multiput(687.00,521.93)(0.874,-0.485){11}{\rule{0.786pt}{0.117pt}}
\multiput(687.00,522.17)(10.369,-7.000){2}{\rule{0.393pt}{0.400pt}}
\multiput(699.00,514.93)(0.874,-0.485){11}{\rule{0.786pt}{0.117pt}}
\multiput(699.00,515.17)(10.369,-7.000){2}{\rule{0.393pt}{0.400pt}}
\multiput(711.00,507.93)(0.950,-0.485){11}{\rule{0.843pt}{0.117pt}}
\multiput(711.00,508.17)(11.251,-7.000){2}{\rule{0.421pt}{0.400pt}}
\multiput(724.00,500.93)(0.758,-0.488){13}{\rule{0.700pt}{0.117pt}}
\multiput(724.00,501.17)(10.547,-8.000){2}{\rule{0.350pt}{0.400pt}}
\multiput(736.00,492.93)(0.874,-0.485){11}{\rule{0.786pt}{0.117pt}}
\multiput(736.00,493.17)(10.369,-7.000){2}{\rule{0.393pt}{0.400pt}}
\multiput(748.00,485.93)(0.874,-0.485){11}{\rule{0.786pt}{0.117pt}}
\multiput(748.00,486.17)(10.369,-7.000){2}{\rule{0.393pt}{0.400pt}}
\multiput(760.00,478.93)(1.123,-0.482){9}{\rule{0.967pt}{0.116pt}}
\multiput(760.00,479.17)(10.994,-6.000){2}{\rule{0.483pt}{0.400pt}}
\multiput(773.00,472.93)(0.874,-0.485){11}{\rule{0.786pt}{0.117pt}}
\multiput(773.00,473.17)(10.369,-7.000){2}{\rule{0.393pt}{0.400pt}}
\multiput(785.00,465.93)(0.874,-0.485){11}{\rule{0.786pt}{0.117pt}}
\multiput(785.00,466.17)(10.369,-7.000){2}{\rule{0.393pt}{0.400pt}}
\multiput(797.00,458.93)(0.950,-0.485){11}{\rule{0.843pt}{0.117pt}}
\multiput(797.00,459.17)(11.251,-7.000){2}{\rule{0.421pt}{0.400pt}}
\multiput(810.00,451.93)(1.033,-0.482){9}{\rule{0.900pt}{0.116pt}}
\multiput(810.00,452.17)(10.132,-6.000){2}{\rule{0.450pt}{0.400pt}}
\multiput(822.00,445.93)(0.874,-0.485){11}{\rule{0.786pt}{0.117pt}}
\multiput(822.00,446.17)(10.369,-7.000){2}{\rule{0.393pt}{0.400pt}}
\multiput(834.00,438.93)(0.874,-0.485){11}{\rule{0.786pt}{0.117pt}}
\multiput(834.00,439.17)(10.369,-7.000){2}{\rule{0.393pt}{0.400pt}}
\multiput(846.00,431.93)(1.123,-0.482){9}{\rule{0.967pt}{0.116pt}}
\multiput(846.00,432.17)(10.994,-6.000){2}{\rule{0.483pt}{0.400pt}}
\multiput(859.00,425.93)(1.033,-0.482){9}{\rule{0.900pt}{0.116pt}}
\multiput(859.00,426.17)(10.132,-6.000){2}{\rule{0.450pt}{0.400pt}}
\multiput(871.00,419.93)(0.874,-0.485){11}{\rule{0.786pt}{0.117pt}}
\multiput(871.00,420.17)(10.369,-7.000){2}{\rule{0.393pt}{0.400pt}}
\multiput(883.00,412.93)(1.123,-0.482){9}{\rule{0.967pt}{0.116pt}}
\multiput(883.00,413.17)(10.994,-6.000){2}{\rule{0.483pt}{0.400pt}}
\multiput(896.00,406.93)(1.033,-0.482){9}{\rule{0.900pt}{0.116pt}}
\multiput(896.00,407.17)(10.132,-6.000){2}{\rule{0.450pt}{0.400pt}}
\multiput(908.00,400.93)(1.033,-0.482){9}{\rule{0.900pt}{0.116pt}}
\multiput(908.00,401.17)(10.132,-6.000){2}{\rule{0.450pt}{0.400pt}}
\multiput(920.00,394.93)(1.033,-0.482){9}{\rule{0.900pt}{0.116pt}}
\multiput(920.00,395.17)(10.132,-6.000){2}{\rule{0.450pt}{0.400pt}}
\multiput(932.00,388.93)(1.123,-0.482){9}{\rule{0.967pt}{0.116pt}}
\multiput(932.00,389.17)(10.994,-6.000){2}{\rule{0.483pt}{0.400pt}}
\multiput(945.00,382.93)(1.033,-0.482){9}{\rule{0.900pt}{0.116pt}}
\multiput(945.00,383.17)(10.132,-6.000){2}{\rule{0.450pt}{0.400pt}}
\multiput(957.00,376.93)(1.033,-0.482){9}{\rule{0.900pt}{0.116pt}}
\multiput(957.00,377.17)(10.132,-6.000){2}{\rule{0.450pt}{0.400pt}}
\multiput(969.00,370.93)(1.123,-0.482){9}{\rule{0.967pt}{0.116pt}}
\multiput(969.00,371.17)(10.994,-6.000){2}{\rule{0.483pt}{0.400pt}}
\multiput(982.00,364.93)(1.033,-0.482){9}{\rule{0.900pt}{0.116pt}}
\multiput(982.00,365.17)(10.132,-6.000){2}{\rule{0.450pt}{0.400pt}}
\multiput(994.00,358.93)(1.033,-0.482){9}{\rule{0.900pt}{0.116pt}}
\multiput(994.00,359.17)(10.132,-6.000){2}{\rule{0.450pt}{0.400pt}}
\multiput(1006.00,352.93)(1.267,-0.477){7}{\rule{1.060pt}{0.115pt}}
\multiput(1006.00,353.17)(9.800,-5.000){2}{\rule{0.530pt}{0.400pt}}
\multiput(1018.00,347.93)(1.123,-0.482){9}{\rule{0.967pt}{0.116pt}}
\multiput(1018.00,348.17)(10.994,-6.000){2}{\rule{0.483pt}{0.400pt}}
\multiput(1031.00,341.93)(1.267,-0.477){7}{\rule{1.060pt}{0.115pt}}
\multiput(1031.00,342.17)(9.800,-5.000){2}{\rule{0.530pt}{0.400pt}}
\multiput(1043.00,336.93)(1.033,-0.482){9}{\rule{0.900pt}{0.116pt}}
\multiput(1043.00,337.17)(10.132,-6.000){2}{\rule{0.450pt}{0.400pt}}
\multiput(1055.00,330.93)(1.378,-0.477){7}{\rule{1.140pt}{0.115pt}}
\multiput(1055.00,331.17)(10.634,-5.000){2}{\rule{0.570pt}{0.400pt}}
\multiput(1068.00,325.93)(1.033,-0.482){9}{\rule{0.900pt}{0.116pt}}
\multiput(1068.00,326.17)(10.132,-6.000){2}{\rule{0.450pt}{0.400pt}}
\multiput(1080.00,319.93)(1.267,-0.477){7}{\rule{1.060pt}{0.115pt}}
\multiput(1080.00,320.17)(9.800,-5.000){2}{\rule{0.530pt}{0.400pt}}
\multiput(1092.00,314.93)(1.267,-0.477){7}{\rule{1.060pt}{0.115pt}}
\multiput(1092.00,315.17)(9.800,-5.000){2}{\rule{0.530pt}{0.400pt}}
\multiput(1104.00,309.93)(1.378,-0.477){7}{\rule{1.140pt}{0.115pt}}
\multiput(1104.00,310.17)(10.634,-5.000){2}{\rule{0.570pt}{0.400pt}}
\multiput(1117.00,304.93)(1.033,-0.482){9}{\rule{0.900pt}{0.116pt}}
\multiput(1117.00,305.17)(10.132,-6.000){2}{\rule{0.450pt}{0.400pt}}
\multiput(1129.00,298.93)(1.267,-0.477){7}{\rule{1.060pt}{0.115pt}}
\multiput(1129.00,299.17)(9.800,-5.000){2}{\rule{0.530pt}{0.400pt}}
\multiput(1141.00,293.93)(1.267,-0.477){7}{\rule{1.060pt}{0.115pt}}
\multiput(1141.00,294.17)(9.800,-5.000){2}{\rule{0.530pt}{0.400pt}}
\multiput(1153.00,288.93)(1.378,-0.477){7}{\rule{1.140pt}{0.115pt}}
\multiput(1153.00,289.17)(10.634,-5.000){2}{\rule{0.570pt}{0.400pt}}
\multiput(1166.00,283.93)(1.267,-0.477){7}{\rule{1.060pt}{0.115pt}}
\multiput(1166.00,284.17)(9.800,-5.000){2}{\rule{0.530pt}{0.400pt}}
\multiput(1178.00,278.93)(1.267,-0.477){7}{\rule{1.060pt}{0.115pt}}
\multiput(1178.00,279.17)(9.800,-5.000){2}{\rule{0.530pt}{0.400pt}}
\multiput(1190.00,273.93)(1.378,-0.477){7}{\rule{1.140pt}{0.115pt}}
\multiput(1190.00,274.17)(10.634,-5.000){2}{\rule{0.570pt}{0.400pt}}
\multiput(1203.00,268.94)(1.651,-0.468){5}{\rule{1.300pt}{0.113pt}}
\multiput(1203.00,269.17)(9.302,-4.000){2}{\rule{0.650pt}{0.400pt}}
\multiput(1215.00,264.93)(1.267,-0.477){7}{\rule{1.060pt}{0.115pt}}
\multiput(1215.00,265.17)(9.800,-5.000){2}{\rule{0.530pt}{0.400pt}}
\multiput(1227.00,259.93)(1.267,-0.477){7}{\rule{1.060pt}{0.115pt}}
\multiput(1227.00,260.17)(9.800,-5.000){2}{\rule{0.530pt}{0.400pt}}
\multiput(1239.00,254.93)(1.378,-0.477){7}{\rule{1.140pt}{0.115pt}}
\multiput(1239.00,255.17)(10.634,-5.000){2}{\rule{0.570pt}{0.400pt}}
\multiput(1252.00,249.94)(1.651,-0.468){5}{\rule{1.300pt}{0.113pt}}
\multiput(1252.00,250.17)(9.302,-4.000){2}{\rule{0.650pt}{0.400pt}}
\multiput(1264.00,245.93)(1.267,-0.477){7}{\rule{1.060pt}{0.115pt}}
\multiput(1264.00,246.17)(9.800,-5.000){2}{\rule{0.530pt}{0.400pt}}
\multiput(1276.00,240.94)(1.797,-0.468){5}{\rule{1.400pt}{0.113pt}}
\multiput(1276.00,241.17)(10.094,-4.000){2}{\rule{0.700pt}{0.400pt}}
\multiput(1289.00,236.93)(1.267,-0.477){7}{\rule{1.060pt}{0.115pt}}
\multiput(1289.00,237.17)(9.800,-5.000){2}{\rule{0.530pt}{0.400pt}}
\multiput(1301.00,231.94)(1.651,-0.468){5}{\rule{1.300pt}{0.113pt}}
\multiput(1301.00,232.17)(9.302,-4.000){2}{\rule{0.650pt}{0.400pt}}
\multiput(1313.00,227.93)(1.267,-0.477){7}{\rule{1.060pt}{0.115pt}}
\multiput(1313.00,228.17)(9.800,-5.000){2}{\rule{0.530pt}{0.400pt}}
\multiput(1325.00,222.94)(1.797,-0.468){5}{\rule{1.400pt}{0.113pt}}
\multiput(1325.00,223.17)(10.094,-4.000){2}{\rule{0.700pt}{0.400pt}}
\multiput(1338.00,218.94)(1.651,-0.468){5}{\rule{1.300pt}{0.113pt}}
\multiput(1338.00,219.17)(9.302,-4.000){2}{\rule{0.650pt}{0.400pt}}
\multiput(1350.00,214.93)(1.267,-0.477){7}{\rule{1.060pt}{0.115pt}}
\multiput(1350.00,215.17)(9.800,-5.000){2}{\rule{0.530pt}{0.400pt}}
\multiput(1362.00,209.94)(1.797,-0.468){5}{\rule{1.400pt}{0.113pt}}
\multiput(1362.00,210.17)(10.094,-4.000){2}{\rule{0.700pt}{0.400pt}}
\multiput(1375.00,205.94)(1.651,-0.468){5}{\rule{1.300pt}{0.113pt}}
\multiput(1375.00,206.17)(9.302,-4.000){2}{\rule{0.650pt}{0.400pt}}
\multiput(1387.00,201.94)(1.651,-0.468){5}{\rule{1.300pt}{0.113pt}}
\multiput(1387.00,202.17)(9.302,-4.000){2}{\rule{0.650pt}{0.400pt}}
\multiput(1399.00,197.94)(1.651,-0.468){5}{\rule{1.300pt}{0.113pt}}
\multiput(1399.00,198.17)(9.302,-4.000){2}{\rule{0.650pt}{0.400pt}}
\multiput(1411.00,193.94)(1.797,-0.468){5}{\rule{1.400pt}{0.113pt}}
\multiput(1411.00,194.17)(10.094,-4.000){2}{\rule{0.700pt}{0.400pt}}
\multiput(1424.00,189.93)(1.267,-0.477){7}{\rule{1.060pt}{0.115pt}}
\multiput(1424.00,190.17)(9.800,-5.000){2}{\rule{0.530pt}{0.400pt}}
\sbox{\plotpoint}{\rule[-0.500pt]{1.000pt}{1.000pt}}%
\put(1306,767){\makebox(0,0)[r]{$\eta=0.8$}}
\multiput(1328,767)(20.756,0.000){4}{\usebox{\plotpoint}}
\put(1394,767){\usebox{\plotpoint}}
\put(220,685){\usebox{\plotpoint}}
\put(220.00,685.00){\usebox{\plotpoint}}
\put(236.73,672.72){\usebox{\plotpoint}}
\put(253.90,661.06){\usebox{\plotpoint}}
\multiput(257,659)(16.604,-12.453){0}{\usebox{\plotpoint}}
\put(270.63,648.78){\usebox{\plotpoint}}
\put(287.63,636.92){\usebox{\plotpoint}}
\put(304.62,625.03){\usebox{\plotpoint}}
\multiput(306,624)(17.270,-11.513){0}{\usebox{\plotpoint}}
\put(321.93,613.58){\usebox{\plotpoint}}
\put(339.41,602.40){\usebox{\plotpoint}}
\multiput(343,600)(16.604,-12.453){0}{\usebox{\plotpoint}}
\put(356.20,590.20){\usebox{\plotpoint}}
\put(373.62,578.93){\usebox{\plotpoint}}
\put(391.46,568.32){\usebox{\plotpoint}}
\multiput(392,568)(17.270,-11.513){0}{\usebox{\plotpoint}}
\put(408.86,557.01){\usebox{\plotpoint}}
\put(426.67,546.36){\usebox{\plotpoint}}
\multiput(429,545)(17.270,-11.513){0}{\usebox{\plotpoint}}
\put(444.02,534.98){\usebox{\plotpoint}}
\put(461.78,524.27){\usebox{\plotpoint}}
\multiput(466,522)(17.928,-10.458){0}{\usebox{\plotpoint}}
\put(479.72,513.85){\usebox{\plotpoint}}
\put(497.40,503.02){\usebox{\plotpoint}}
\multiput(503,500)(17.928,-10.458){0}{\usebox{\plotpoint}}
\put(515.43,492.75){\usebox{\plotpoint}}
\put(533.36,482.29){\usebox{\plotpoint}}
\put(551.52,472.26){\usebox{\plotpoint}}
\multiput(552,472)(17.928,-10.458){0}{\usebox{\plotpoint}}
\put(569.65,462.17){\usebox{\plotpoint}}
\put(587.80,452.12){\usebox{\plotpoint}}
\multiput(588,452)(18.275,-9.840){0}{\usebox{\plotpoint}}
\put(606.15,442.42){\usebox{\plotpoint}}
\put(624.31,432.40){\usebox{\plotpoint}}
\multiput(625,432)(18.845,-8.698){0}{\usebox{\plotpoint}}
\put(642.88,423.16){\usebox{\plotpoint}}
\put(661.19,413.41){\usebox{\plotpoint}}
\multiput(662,413)(18.564,-9.282){0}{\usebox{\plotpoint}}
\put(679.84,404.31){\usebox{\plotpoint}}
\put(698.51,395.25){\usebox{\plotpoint}}
\multiput(699,395)(18.564,-9.282){0}{\usebox{\plotpoint}}
\put(717.16,386.15){\usebox{\plotpoint}}
\put(735.83,377.08){\usebox{\plotpoint}}
\multiput(736,377)(18.564,-9.282){0}{\usebox{\plotpoint}}
\put(754.40,367.80){\usebox{\plotpoint}}
\multiput(760,365)(18.845,-8.698){0}{\usebox{\plotpoint}}
\put(773.15,358.92){\usebox{\plotpoint}}
\put(791.93,350.11){\usebox{\plotpoint}}
\multiput(797,348)(18.845,-8.698){0}{\usebox{\plotpoint}}
\put(810.88,341.64){\usebox{\plotpoint}}
\put(829.78,333.11){\usebox{\plotpoint}}
\multiput(834,331)(19.159,-7.983){0}{\usebox{\plotpoint}}
\put(848.76,324.72){\usebox{\plotpoint}}
\put(867.75,316.35){\usebox{\plotpoint}}
\multiput(871,315)(19.159,-7.983){0}{\usebox{\plotpoint}}
\put(886.95,308.48){\usebox{\plotpoint}}
\put(905.90,300.05){\usebox{\plotpoint}}
\multiput(908,299)(19.159,-7.983){0}{\usebox{\plotpoint}}
\put(924.99,291.92){\usebox{\plotpoint}}
\put(944.28,284.28){\usebox{\plotpoint}}
\multiput(945,284)(19.159,-7.983){0}{\usebox{\plotpoint}}
\put(963.45,276.31){\usebox{\plotpoint}}
\multiput(969,274)(19.372,-7.451){0}{\usebox{\plotpoint}}
\put(982.77,268.74){\usebox{\plotpoint}}
\put(1002.23,261.57){\usebox{\plotpoint}}
\multiput(1006,260)(19.159,-7.983){0}{\usebox{\plotpoint}}
\put(1021.51,253.92){\usebox{\plotpoint}}
\put(1040.99,246.84){\usebox{\plotpoint}}
\multiput(1043,246)(19.159,-7.983){0}{\usebox{\plotpoint}}
\put(1060.34,239.36){\usebox{\plotpoint}}
\put(1079.76,232.10){\usebox{\plotpoint}}
\multiput(1080,232)(19.690,-6.563){0}{\usebox{\plotpoint}}
\put(1099.24,224.98){\usebox{\plotpoint}}
\multiput(1104,223)(19.838,-6.104){0}{\usebox{\plotpoint}}
\put(1118.90,218.37){\usebox{\plotpoint}}
\put(1138.59,211.80){\usebox{\plotpoint}}
\multiput(1141,211)(19.159,-7.983){0}{\usebox{\plotpoint}}
\put(1157.98,204.47){\usebox{\plotpoint}}
\put(1177.73,198.09){\usebox{\plotpoint}}
\multiput(1178,198)(19.690,-6.563){0}{\usebox{\plotpoint}}
\put(1197.48,191.70){\usebox{\plotpoint}}
\multiput(1203,190)(19.690,-6.563){0}{\usebox{\plotpoint}}
\put(1217.21,185.26){\usebox{\plotpoint}}
\put(1236.90,178.70){\usebox{\plotpoint}}
\multiput(1239,178)(19.838,-6.104){0}{\usebox{\plotpoint}}
\put(1256.68,172.44){\usebox{\plotpoint}}
\multiput(1264,170)(19.690,-6.563){0}{\usebox{\plotpoint}}
\put(1276.38,165.88){\usebox{\plotpoint}}
\put(1296.32,160.17){\usebox{\plotpoint}}
\multiput(1301,159)(19.690,-6.563){0}{\usebox{\plotpoint}}
\put(1316.12,153.96){\usebox{\plotpoint}}
\put(1336.10,148.44){\usebox{\plotpoint}}
\multiput(1338,148)(19.690,-6.563){0}{\usebox{\plotpoint}}
\put(1355.97,142.51){\usebox{\plotpoint}}
\multiput(1362,141)(19.838,-6.104){0}{\usebox{\plotpoint}}
\put(1375.89,136.70){\usebox{\plotpoint}}
\put(1395.78,130.81){\usebox{\plotpoint}}
\multiput(1399,130)(20.136,-5.034){0}{\usebox{\plotpoint}}
\put(1415.84,125.51){\usebox{\plotpoint}}
\put(1435.85,120.04){\usebox{\plotpoint}}
\put(1436,120){\usebox{\plotpoint}}
\end{picture}

%% file: fig5.tex
\setlength{\unitlength}{0.240900pt}
\ifx\plotpoint\undefined\newsavebox{\plotpoint}\fi
\sbox{\plotpoint}{\rule[-0.200pt]{0.400pt}{0.400pt}}%
\begin{picture}(1500,900)(0,0)
\font\gnuplot=cmr10 at 10pt
\gnuplot
\sbox{\plotpoint}{\rule[-0.200pt]{0.400pt}{0.400pt}}%
\put(220.0,113.0){\rule[-0.200pt]{292.934pt}{0.400pt}}
\put(220.0,113.0){\rule[-0.200pt]{4.818pt}{0.400pt}}
\put(198,113){\makebox(0,0)[r]{0}}
\put(1416.0,113.0){\rule[-0.200pt]{4.818pt}{0.400pt}}
\put(220.0,189.0){\rule[-0.200pt]{4.818pt}{0.400pt}}
\put(198,189){\makebox(0,0)[r]{0.1}}
\put(1416.0,189.0){\rule[-0.200pt]{4.818pt}{0.400pt}}
\put(220.0,266.0){\rule[-0.200pt]{4.818pt}{0.400pt}}
\put(198,266){\makebox(0,0)[r]{0.2}}
\put(1416.0,266.0){\rule[-0.200pt]{4.818pt}{0.400pt}}
\put(220.0,342.0){\rule[-0.200pt]{4.818pt}{0.400pt}}
\put(198,342){\makebox(0,0)[r]{0.3}}
\put(1416.0,342.0){\rule[-0.200pt]{4.818pt}{0.400pt}}
\put(220.0,419.0){\rule[-0.200pt]{4.818pt}{0.400pt}}
\put(198,419){\makebox(0,0)[r]{0.4}}
\put(1416.0,419.0){\rule[-0.200pt]{4.818pt}{0.400pt}}
\put(220.0,495.0){\rule[-0.200pt]{4.818pt}{0.400pt}}
\put(198,495){\makebox(0,0)[r]{0.5}}
\put(1416.0,495.0){\rule[-0.200pt]{4.818pt}{0.400pt}}
\put(220.0,571.0){\rule[-0.200pt]{4.818pt}{0.400pt}}
\put(198,571){\makebox(0,0)[r]{0.6}}
\put(1416.0,571.0){\rule[-0.200pt]{4.818pt}{0.400pt}}
\put(220.0,648.0){\rule[-0.200pt]{4.818pt}{0.400pt}}
\put(198,648){\makebox(0,0)[r]{0.7}}
\put(1416.0,648.0){\rule[-0.200pt]{4.818pt}{0.400pt}}
\put(220.0,724.0){\rule[-0.200pt]{4.818pt}{0.400pt}}
\put(198,724){\makebox(0,0)[r]{0.8}}
\put(1416.0,724.0){\rule[-0.200pt]{4.818pt}{0.400pt}}
\put(220.0,801.0){\rule[-0.200pt]{4.818pt}{0.400pt}}
\put(198,801){\makebox(0,0)[r]{0.9}}
\put(1416.0,801.0){\rule[-0.200pt]{4.818pt}{0.400pt}}
\put(220.0,877.0){\rule[-0.200pt]{4.818pt}{0.400pt}}
\put(198,877){\makebox(0,0)[r]{1}}
\put(1416.0,877.0){\rule[-0.200pt]{4.818pt}{0.400pt}}
\put(331.0,113.0){\rule[-0.200pt]{0.400pt}{4.818pt}}
\put(331,68){\makebox(0,0){50}}
\put(331.0,857.0){\rule[-0.200pt]{0.400pt}{4.818pt}}
\put(453.0,113.0){\rule[-0.200pt]{0.400pt}{4.818pt}}
\put(453,68){\makebox(0,0){100}}
\put(453.0,857.0){\rule[-0.200pt]{0.400pt}{4.818pt}}
\put(576.0,113.0){\rule[-0.200pt]{0.400pt}{4.818pt}}
\put(576,68){\makebox(0,0){150}}
\put(576.0,857.0){\rule[-0.200pt]{0.400pt}{4.818pt}}
\put(699.0,113.0){\rule[-0.200pt]{0.400pt}{4.818pt}}
\put(699,68){\makebox(0,0){200}}
\put(699.0,857.0){\rule[-0.200pt]{0.400pt}{4.818pt}}
\put(822.0,113.0){\rule[-0.200pt]{0.400pt}{4.818pt}}
\put(822,68){\makebox(0,0){250}}
\put(822.0,857.0){\rule[-0.200pt]{0.400pt}{4.818pt}}
\put(945.0,113.0){\rule[-0.200pt]{0.400pt}{4.818pt}}
\put(945,68){\makebox(0,0){300}}
\put(945.0,857.0){\rule[-0.200pt]{0.400pt}{4.818pt}}
\put(1068.0,113.0){\rule[-0.200pt]{0.400pt}{4.818pt}}
\put(1068,68){\makebox(0,0){350}}
\put(1068.0,857.0){\rule[-0.200pt]{0.400pt}{4.818pt}}
\put(1190.0,113.0){\rule[-0.200pt]{0.400pt}{4.818pt}}
\put(1190,68){\makebox(0,0){400}}
\put(1190.0,857.0){\rule[-0.200pt]{0.400pt}{4.818pt}}
\put(1313.0,113.0){\rule[-0.200pt]{0.400pt}{4.818pt}}
\put(1313,68){\makebox(0,0){450}}
\put(1313.0,857.0){\rule[-0.200pt]{0.400pt}{4.818pt}}
\put(1436.0,113.0){\rule[-0.200pt]{0.400pt}{4.818pt}}
\put(1436,68){\makebox(0,0){500}}
\put(1436.0,857.0){\rule[-0.200pt]{0.400pt}{4.818pt}}
\put(220.0,113.0){\rule[-0.200pt]{292.934pt}{0.400pt}}
\put(1436.0,113.0){\rule[-0.200pt]{0.400pt}{184.048pt}}
\put(220.0,877.0){\rule[-0.200pt]{292.934pt}{0.400pt}}
\put(45,495){\makebox(0,0){$E$}}
\put(828,23){\makebox(0,0){$t~[1/g]$}}
\put(220.0,113.0){\rule[-0.200pt]{0.400pt}{184.048pt}}
\put(220,817){\usebox{\plotpoint}}
\multiput(220.58,811.33)(0.492,-1.616){21}{\rule{0.119pt}{1.367pt}}
\multiput(219.17,814.16)(12.000,-35.163){2}{\rule{0.400pt}{0.683pt}}
\multiput(232.58,774.50)(0.493,-1.250){23}{\rule{0.119pt}{1.085pt}}
\multiput(231.17,776.75)(13.000,-29.749){2}{\rule{0.400pt}{0.542pt}}
\multiput(245.58,742.71)(0.492,-1.186){21}{\rule{0.119pt}{1.033pt}}
\multiput(244.17,744.86)(12.000,-25.855){2}{\rule{0.400pt}{0.517pt}}
\multiput(257.58,714.99)(0.492,-1.099){21}{\rule{0.119pt}{0.967pt}}
\multiput(256.17,716.99)(12.000,-23.994){2}{\rule{0.400pt}{0.483pt}}
\multiput(269.58,689.40)(0.492,-0.970){21}{\rule{0.119pt}{0.867pt}}
\multiput(268.17,691.20)(12.000,-21.201){2}{\rule{0.400pt}{0.433pt}}
\multiput(281.58,666.77)(0.493,-0.853){23}{\rule{0.119pt}{0.777pt}}
\multiput(280.17,668.39)(13.000,-20.387){2}{\rule{0.400pt}{0.388pt}}
\multiput(294.58,644.68)(0.492,-0.884){21}{\rule{0.119pt}{0.800pt}}
\multiput(293.17,646.34)(12.000,-19.340){2}{\rule{0.400pt}{0.400pt}}
\multiput(306.58,623.96)(0.492,-0.798){21}{\rule{0.119pt}{0.733pt}}
\multiput(305.17,625.48)(12.000,-17.478){2}{\rule{0.400pt}{0.367pt}}
\multiput(318.58,605.41)(0.493,-0.655){23}{\rule{0.119pt}{0.623pt}}
\multiput(317.17,606.71)(13.000,-15.707){2}{\rule{0.400pt}{0.312pt}}
\multiput(331.58,588.23)(0.492,-0.712){21}{\rule{0.119pt}{0.667pt}}
\multiput(330.17,589.62)(12.000,-15.616){2}{\rule{0.400pt}{0.333pt}}
\multiput(343.58,571.37)(0.492,-0.669){21}{\rule{0.119pt}{0.633pt}}
\multiput(342.17,572.69)(12.000,-14.685){2}{\rule{0.400pt}{0.317pt}}
\multiput(355.58,555.51)(0.492,-0.625){21}{\rule{0.119pt}{0.600pt}}
\multiput(354.17,556.75)(12.000,-13.755){2}{\rule{0.400pt}{0.300pt}}
\multiput(367.58,540.80)(0.493,-0.536){23}{\rule{0.119pt}{0.531pt}}
\multiput(366.17,541.90)(13.000,-12.898){2}{\rule{0.400pt}{0.265pt}}
\multiput(380.58,526.65)(0.492,-0.582){21}{\rule{0.119pt}{0.567pt}}
\multiput(379.17,527.82)(12.000,-12.824){2}{\rule{0.400pt}{0.283pt}}
\multiput(392.58,512.79)(0.492,-0.539){21}{\rule{0.119pt}{0.533pt}}
\multiput(391.17,513.89)(12.000,-11.893){2}{\rule{0.400pt}{0.267pt}}
\multiput(404.00,500.92)(0.539,-0.492){21}{\rule{0.533pt}{0.119pt}}
\multiput(404.00,501.17)(11.893,-12.000){2}{\rule{0.267pt}{0.400pt}}
\multiput(417.00,488.92)(0.496,-0.492){21}{\rule{0.500pt}{0.119pt}}
\multiput(417.00,489.17)(10.962,-12.000){2}{\rule{0.250pt}{0.400pt}}
\multiput(429.00,476.92)(0.543,-0.492){19}{\rule{0.536pt}{0.118pt}}
\multiput(429.00,477.17)(10.887,-11.000){2}{\rule{0.268pt}{0.400pt}}
\multiput(441.00,465.92)(0.543,-0.492){19}{\rule{0.536pt}{0.118pt}}
\multiput(441.00,466.17)(10.887,-11.000){2}{\rule{0.268pt}{0.400pt}}
\multiput(453.00,454.92)(0.590,-0.492){19}{\rule{0.573pt}{0.118pt}}
\multiput(453.00,455.17)(11.811,-11.000){2}{\rule{0.286pt}{0.400pt}}
\multiput(466.00,443.92)(0.600,-0.491){17}{\rule{0.580pt}{0.118pt}}
\multiput(466.00,444.17)(10.796,-10.000){2}{\rule{0.290pt}{0.400pt}}
\multiput(478.00,433.93)(0.669,-0.489){15}{\rule{0.633pt}{0.118pt}}
\multiput(478.00,434.17)(10.685,-9.000){2}{\rule{0.317pt}{0.400pt}}
\multiput(490.00,424.93)(0.728,-0.489){15}{\rule{0.678pt}{0.118pt}}
\multiput(490.00,425.17)(11.593,-9.000){2}{\rule{0.339pt}{0.400pt}}
\multiput(503.00,415.93)(0.669,-0.489){15}{\rule{0.633pt}{0.118pt}}
\multiput(503.00,416.17)(10.685,-9.000){2}{\rule{0.317pt}{0.400pt}}
\multiput(515.00,406.93)(0.669,-0.489){15}{\rule{0.633pt}{0.118pt}}
\multiput(515.00,407.17)(10.685,-9.000){2}{\rule{0.317pt}{0.400pt}}
\multiput(527.00,397.93)(0.758,-0.488){13}{\rule{0.700pt}{0.117pt}}
\multiput(527.00,398.17)(10.547,-8.000){2}{\rule{0.350pt}{0.400pt}}
\multiput(539.00,389.93)(0.824,-0.488){13}{\rule{0.750pt}{0.117pt}}
\multiput(539.00,390.17)(11.443,-8.000){2}{\rule{0.375pt}{0.400pt}}
\multiput(552.00,381.93)(0.874,-0.485){11}{\rule{0.786pt}{0.117pt}}
\multiput(552.00,382.17)(10.369,-7.000){2}{\rule{0.393pt}{0.400pt}}
\multiput(564.00,374.93)(0.758,-0.488){13}{\rule{0.700pt}{0.117pt}}
\multiput(564.00,375.17)(10.547,-8.000){2}{\rule{0.350pt}{0.400pt}}
\multiput(576.00,366.93)(0.874,-0.485){11}{\rule{0.786pt}{0.117pt}}
\multiput(576.00,367.17)(10.369,-7.000){2}{\rule{0.393pt}{0.400pt}}
\multiput(588.00,359.93)(0.950,-0.485){11}{\rule{0.843pt}{0.117pt}}
\multiput(588.00,360.17)(11.251,-7.000){2}{\rule{0.421pt}{0.400pt}}
\multiput(601.00,352.93)(1.033,-0.482){9}{\rule{0.900pt}{0.116pt}}
\multiput(601.00,353.17)(10.132,-6.000){2}{\rule{0.450pt}{0.400pt}}
\multiput(613.00,346.93)(0.874,-0.485){11}{\rule{0.786pt}{0.117pt}}
\multiput(613.00,347.17)(10.369,-7.000){2}{\rule{0.393pt}{0.400pt}}
\multiput(625.00,339.93)(1.123,-0.482){9}{\rule{0.967pt}{0.116pt}}
\multiput(625.00,340.17)(10.994,-6.000){2}{\rule{0.483pt}{0.400pt}}
\multiput(638.00,333.93)(1.033,-0.482){9}{\rule{0.900pt}{0.116pt}}
\multiput(638.00,334.17)(10.132,-6.000){2}{\rule{0.450pt}{0.400pt}}
\multiput(650.00,327.93)(1.267,-0.477){7}{\rule{1.060pt}{0.115pt}}
\multiput(650.00,328.17)(9.800,-5.000){2}{\rule{0.530pt}{0.400pt}}
\multiput(662.00,322.93)(1.033,-0.482){9}{\rule{0.900pt}{0.116pt}}
\multiput(662.00,323.17)(10.132,-6.000){2}{\rule{0.450pt}{0.400pt}}
\multiput(674.00,316.93)(1.378,-0.477){7}{\rule{1.140pt}{0.115pt}}
\multiput(674.00,317.17)(10.634,-5.000){2}{\rule{0.570pt}{0.400pt}}
\multiput(687.00,311.93)(1.033,-0.482){9}{\rule{0.900pt}{0.116pt}}
\multiput(687.00,312.17)(10.132,-6.000){2}{\rule{0.450pt}{0.400pt}}
\multiput(699.00,305.93)(1.267,-0.477){7}{\rule{1.060pt}{0.115pt}}
\multiput(699.00,306.17)(9.800,-5.000){2}{\rule{0.530pt}{0.400pt}}
\multiput(711.00,300.93)(1.378,-0.477){7}{\rule{1.140pt}{0.115pt}}
\multiput(711.00,301.17)(10.634,-5.000){2}{\rule{0.570pt}{0.400pt}}
\multiput(724.00,295.94)(1.651,-0.468){5}{\rule{1.300pt}{0.113pt}}
\multiput(724.00,296.17)(9.302,-4.000){2}{\rule{0.650pt}{0.400pt}}
\multiput(736.00,291.93)(1.267,-0.477){7}{\rule{1.060pt}{0.115pt}}
\multiput(736.00,292.17)(9.800,-5.000){2}{\rule{0.530pt}{0.400pt}}
\multiput(748.00,286.93)(1.267,-0.477){7}{\rule{1.060pt}{0.115pt}}
\multiput(748.00,287.17)(9.800,-5.000){2}{\rule{0.530pt}{0.400pt}}
\multiput(760.00,281.94)(1.797,-0.468){5}{\rule{1.400pt}{0.113pt}}
\multiput(760.00,282.17)(10.094,-4.000){2}{\rule{0.700pt}{0.400pt}}
\multiput(773.00,277.94)(1.651,-0.468){5}{\rule{1.300pt}{0.113pt}}
\multiput(773.00,278.17)(9.302,-4.000){2}{\rule{0.650pt}{0.400pt}}
\multiput(785.00,273.94)(1.651,-0.468){5}{\rule{1.300pt}{0.113pt}}
\multiput(785.00,274.17)(9.302,-4.000){2}{\rule{0.650pt}{0.400pt}}
\multiput(797.00,269.94)(1.797,-0.468){5}{\rule{1.400pt}{0.113pt}}
\multiput(797.00,270.17)(10.094,-4.000){2}{\rule{0.700pt}{0.400pt}}
\multiput(810.00,265.94)(1.651,-0.468){5}{\rule{1.300pt}{0.113pt}}
\multiput(810.00,266.17)(9.302,-4.000){2}{\rule{0.650pt}{0.400pt}}
\multiput(822.00,261.94)(1.651,-0.468){5}{\rule{1.300pt}{0.113pt}}
\multiput(822.00,262.17)(9.302,-4.000){2}{\rule{0.650pt}{0.400pt}}
\multiput(834.00,257.95)(2.472,-0.447){3}{\rule{1.700pt}{0.108pt}}
\multiput(834.00,258.17)(8.472,-3.000){2}{\rule{0.850pt}{0.400pt}}
\multiput(846.00,254.94)(1.797,-0.468){5}{\rule{1.400pt}{0.113pt}}
\multiput(846.00,255.17)(10.094,-4.000){2}{\rule{0.700pt}{0.400pt}}
\multiput(859.00,250.95)(2.472,-0.447){3}{\rule{1.700pt}{0.108pt}}
\multiput(859.00,251.17)(8.472,-3.000){2}{\rule{0.850pt}{0.400pt}}
\multiput(871.00,247.94)(1.651,-0.468){5}{\rule{1.300pt}{0.113pt}}
\multiput(871.00,248.17)(9.302,-4.000){2}{\rule{0.650pt}{0.400pt}}
\multiput(883.00,243.95)(2.695,-0.447){3}{\rule{1.833pt}{0.108pt}}
\multiput(883.00,244.17)(9.195,-3.000){2}{\rule{0.917pt}{0.400pt}}
\multiput(896.00,240.95)(2.472,-0.447){3}{\rule{1.700pt}{0.108pt}}
\multiput(896.00,241.17)(8.472,-3.000){2}{\rule{0.850pt}{0.400pt}}
\multiput(908.00,237.95)(2.472,-0.447){3}{\rule{1.700pt}{0.108pt}}
\multiput(908.00,238.17)(8.472,-3.000){2}{\rule{0.850pt}{0.400pt}}
\multiput(920.00,234.95)(2.472,-0.447){3}{\rule{1.700pt}{0.108pt}}
\multiput(920.00,235.17)(8.472,-3.000){2}{\rule{0.850pt}{0.400pt}}
\multiput(932.00,231.95)(2.695,-0.447){3}{\rule{1.833pt}{0.108pt}}
\multiput(932.00,232.17)(9.195,-3.000){2}{\rule{0.917pt}{0.400pt}}
\multiput(945.00,228.95)(2.472,-0.447){3}{\rule{1.700pt}{0.108pt}}
\multiput(945.00,229.17)(8.472,-3.000){2}{\rule{0.850pt}{0.400pt}}
\multiput(957.00,225.95)(2.472,-0.447){3}{\rule{1.700pt}{0.108pt}}
\multiput(957.00,226.17)(8.472,-3.000){2}{\rule{0.850pt}{0.400pt}}
\put(969,222.17){\rule{2.700pt}{0.400pt}}
\multiput(969.00,223.17)(7.396,-2.000){2}{\rule{1.350pt}{0.400pt}}
\multiput(982.00,220.95)(2.472,-0.447){3}{\rule{1.700pt}{0.108pt}}
\multiput(982.00,221.17)(8.472,-3.000){2}{\rule{0.850pt}{0.400pt}}
\multiput(994.00,217.95)(2.472,-0.447){3}{\rule{1.700pt}{0.108pt}}
\multiput(994.00,218.17)(8.472,-3.000){2}{\rule{0.850pt}{0.400pt}}
\put(1006,214.17){\rule{2.500pt}{0.400pt}}
\multiput(1006.00,215.17)(6.811,-2.000){2}{\rule{1.250pt}{0.400pt}}
\put(1018,212.17){\rule{2.700pt}{0.400pt}}
\multiput(1018.00,213.17)(7.396,-2.000){2}{\rule{1.350pt}{0.400pt}}
\multiput(1031.00,210.95)(2.472,-0.447){3}{\rule{1.700pt}{0.108pt}}
\multiput(1031.00,211.17)(8.472,-3.000){2}{\rule{0.850pt}{0.400pt}}
\put(1043,207.17){\rule{2.500pt}{0.400pt}}
\multiput(1043.00,208.17)(6.811,-2.000){2}{\rule{1.250pt}{0.400pt}}
\put(1055,205.17){\rule{2.700pt}{0.400pt}}
\multiput(1055.00,206.17)(7.396,-2.000){2}{\rule{1.350pt}{0.400pt}}
\put(1068,203.17){\rule{2.500pt}{0.400pt}}
\multiput(1068.00,204.17)(6.811,-2.000){2}{\rule{1.250pt}{0.400pt}}
\put(1080,201.17){\rule{2.500pt}{0.400pt}}
\multiput(1080.00,202.17)(6.811,-2.000){2}{\rule{1.250pt}{0.400pt}}
\multiput(1092.00,199.95)(2.472,-0.447){3}{\rule{1.700pt}{0.108pt}}
\multiput(1092.00,200.17)(8.472,-3.000){2}{\rule{0.850pt}{0.400pt}}
\put(1104,196.17){\rule{2.700pt}{0.400pt}}
\multiput(1104.00,197.17)(7.396,-2.000){2}{\rule{1.350pt}{0.400pt}}
\put(1117,194.67){\rule{2.891pt}{0.400pt}}
\multiput(1117.00,195.17)(6.000,-1.000){2}{\rule{1.445pt}{0.400pt}}
\put(1129,193.17){\rule{2.500pt}{0.400pt}}
\multiput(1129.00,194.17)(6.811,-2.000){2}{\rule{1.250pt}{0.400pt}}
\put(1141,191.17){\rule{2.500pt}{0.400pt}}
\multiput(1141.00,192.17)(6.811,-2.000){2}{\rule{1.250pt}{0.400pt}}
\put(1153,189.17){\rule{2.700pt}{0.400pt}}
\multiput(1153.00,190.17)(7.396,-2.000){2}{\rule{1.350pt}{0.400pt}}
\put(1166,187.17){\rule{2.500pt}{0.400pt}}
\multiput(1166.00,188.17)(6.811,-2.000){2}{\rule{1.250pt}{0.400pt}}
\put(1178,185.67){\rule{2.891pt}{0.400pt}}
\multiput(1178.00,186.17)(6.000,-1.000){2}{\rule{1.445pt}{0.400pt}}
\put(1190,184.17){\rule{2.700pt}{0.400pt}}
\multiput(1190.00,185.17)(7.396,-2.000){2}{\rule{1.350pt}{0.400pt}}
\put(1203,182.17){\rule{2.500pt}{0.400pt}}
\multiput(1203.00,183.17)(6.811,-2.000){2}{\rule{1.250pt}{0.400pt}}
\put(1215,180.67){\rule{2.891pt}{0.400pt}}
\multiput(1215.00,181.17)(6.000,-1.000){2}{\rule{1.445pt}{0.400pt}}
\put(1227,179.17){\rule{2.500pt}{0.400pt}}
\multiput(1227.00,180.17)(6.811,-2.000){2}{\rule{1.250pt}{0.400pt}}
\put(1239,177.67){\rule{3.132pt}{0.400pt}}
\multiput(1239.00,178.17)(6.500,-1.000){2}{\rule{1.566pt}{0.400pt}}
\put(1252,176.17){\rule{2.500pt}{0.400pt}}
\multiput(1252.00,177.17)(6.811,-2.000){2}{\rule{1.250pt}{0.400pt}}
\put(1264,174.67){\rule{2.891pt}{0.400pt}}
\multiput(1264.00,175.17)(6.000,-1.000){2}{\rule{1.445pt}{0.400pt}}
\put(1276,173.17){\rule{2.700pt}{0.400pt}}
\multiput(1276.00,174.17)(7.396,-2.000){2}{\rule{1.350pt}{0.400pt}}
\put(1289,171.67){\rule{2.891pt}{0.400pt}}
\multiput(1289.00,172.17)(6.000,-1.000){2}{\rule{1.445pt}{0.400pt}}
\put(1301,170.67){\rule{2.891pt}{0.400pt}}
\multiput(1301.00,171.17)(6.000,-1.000){2}{\rule{1.445pt}{0.400pt}}
\put(1313,169.17){\rule{2.500pt}{0.400pt}}
\multiput(1313.00,170.17)(6.811,-2.000){2}{\rule{1.250pt}{0.400pt}}
\put(1325,167.67){\rule{3.132pt}{0.400pt}}
\multiput(1325.00,168.17)(6.500,-1.000){2}{\rule{1.566pt}{0.400pt}}
\put(1338,166.67){\rule{2.891pt}{0.400pt}}
\multiput(1338.00,167.17)(6.000,-1.000){2}{\rule{1.445pt}{0.400pt}}
\put(1350,165.67){\rule{2.891pt}{0.400pt}}
\multiput(1350.00,166.17)(6.000,-1.000){2}{\rule{1.445pt}{0.400pt}}
\put(1362,164.17){\rule{2.700pt}{0.400pt}}
\multiput(1362.00,165.17)(7.396,-2.000){2}{\rule{1.350pt}{0.400pt}}
\put(1375,162.67){\rule{2.891pt}{0.400pt}}
\multiput(1375.00,163.17)(6.000,-1.000){2}{\rule{1.445pt}{0.400pt}}
\put(1387,161.67){\rule{2.891pt}{0.400pt}}
\multiput(1387.00,162.17)(6.000,-1.000){2}{\rule{1.445pt}{0.400pt}}
\put(1399,160.67){\rule{2.891pt}{0.400pt}}
\multiput(1399.00,161.17)(6.000,-1.000){2}{\rule{1.445pt}{0.400pt}}
\put(1411,159.67){\rule{3.132pt}{0.400pt}}
\multiput(1411.00,160.17)(6.500,-1.000){2}{\rule{1.566pt}{0.400pt}}
\put(1424,158.67){\rule{2.891pt}{0.400pt}}
\multiput(1424.00,159.17)(6.000,-1.000){2}{\rule{1.445pt}{0.400pt}}
\end{picture}

%% file: published.bbl
\begin{references}

\bibitem{PLE98a} M.B. Plenio and V. Vedral, to appear in Cont. Phys. 1998.
%
\bibitem{epr}
E. Schr{\"o}dinger, Naturwissenschaften {\bf 23}, 807 (1935)
[An English translation appears in {\em Quantum Theory and Measurement},
eds J.A. Wheeler and W.H. Zurek, (Princeton University Press, Princeton,
New Jersey, 1983) pp 152-157]; A. Einstein, B. Podolsky and N. Rosen, Phys. Rev. {\bf 47}, 777 (1935).
%
\bibitem{bell} J.S. Bell, Rev. Mod. Phys. {\bf 38}, 447 (1966).
%
\bibitem{qi1}
R.F. Werner, Phys. Rev. A {\bf 40}, 4277 (1989); N. Gisin, Phys. Lett. A {\bf 210}, 151 (1996).
%
\bibitem{qi2}
A. Peres, Phys. Rev. A {\bf 54}, 2685 (1996);
M. Horodecki, P. Horodecki and R. Horodecki, Phys. Lett. A {\bf 223}, 1 (1997).
%
\bibitem{qi3}
C.H. Bennett, D. DiVincenzo, J. Smolin, and W. Wootters, Phys. Rev. A {\bf 54}, 3824 (1996);
W. Wootters, Phys. Rev. Lett. {\bf 80}, (1998);
V. Vedral, M.B. Plenio, M.A. Rippin, and P.L. Knight, Phys. Rev. Lett. {\bf 78},
2275 (1997), V. Vedral, M.B. Plenio, K. Jacobs, and P.L. Knight, Phys. Rev. A
{\bf 56}, 4452 (1997);
J. Eisert and M.B. Plenio, to appear in J. Mod. Opt. 1998.
%
\bibitem{matpra}
V. Vedral and M. B. Plenio, Phys. Rev. A {\bf 57}, 1619 (1998).
%
\bibitem{qi4}
A. Barenco, Cont. Phys. {\bf 37}, 375 (1996);
V. Vedral and M.B. Plenio, Prog. Quant. Elect. {\bf 22}, 1 (1998).
%
\bibitem{qi4a} C.H. Bennett and G. Brassard, Proceedings of the IEEE International
Conference on
Computers, Systems and Signal Processing, Bangalore, India (IEEE, New York 1984),
pp 175;  A. K. Ekert, Phys. Rev. Lett. {\bf 67 }, 661 (1991).
%
\bibitem{susana}
J. J. Bollinger et al., {Phys. Rev.} A{\bf 54}, R4649 (1996);
S.F. Huelga, C. Macchiavello, T. Pellizari, A.K. Ekert, M.B. Plenio, and J.I. Cirac,
Phys. Rev. Lett. {\bf 79}, 3865 (1997).
%
\bibitem{qi5}
C. Monroe et al.,  Phys. Rev. Lett. {\bf 75 }, 4714 (1995); Q. Turchette et al.,
 Phys. Rev. Lett. {\bf 75 }, 4710 (1995); X. Maitre et al., Phys. Rev. Lett. {\bf 79 }, 769 (1997);
I. Chuang et al., Nature {\bf 393}, 143 (1998); J. Jones, M. Mosca and R. H. Hansen, Nature
{\bf 393}, 344 (1998).
%
\bibitem{rewWineland}
D. W. Wineland et al., to appear in NIST Journal of Research; e-print at quant-ph/9803023
and references therein.
%
\bibitem{haroche}
S. Haroche, M. Brune and J.M. Raimond,
Phil. Trans. of the Royal Soc. of London, A{\bf 355}, 2367 (1997) and references therein.
%
\bibitem{knight}
P.M. Radmore and P.L. Knight, J. Phys. B {\bf 15}, 561 (1982).
%
\bibitem{klaus}
P. Horak and K.M. Gheri, Phys. Rev. A {\bf 53}, R1970 (1996);
K.M. Gheri, P. Horak and H. Ritsch, J. Mod. Opt. {\bf 44}, 605 (1997).
%
\bibitem{carlm}
P. Kochan et al., Phys. Rev. Lett. {\bf 75}, 45 (1995).
%
\bibitem{other} Other probabilistic schemes for the generation of entangled
states of massive particles can be find in
E.S. Fry, T. Walther, and S. Li, Phys. Rev. A {\bf 52}, 4381 (1995)
and E. Hagley et al., Phys. Rev. Lett. {\bf 79}, 1 (1997).
%
\bibitem{prospect}
The influence of the dipole-dipole interaction when the atoms become closer
will be analysed
elsewhere.
%
\bibitem{qj1} C. Cohen-Tannoudji and J. Dalibard, Europhys. Lett. {\bf 1}, 441 (1986);
J. Dalibard, Y. Castin and K. Molmer, Phys. Rev. Lett. {\bf 68}, 580 (1992).
%
\bibitem{qj2} G. C. Hegerfeldt and T. S. Wilser, {\em Proceedings of the II International
Wigner Symposium, Goslar 1991.}, H. D. Doebner, W. Scherer and F. Schroeck Eds.,
(World Scientific, Singapore, 1992).
%
\bibitem{qj3} H. J. Carmichael, {\em An Open Systems Approach to Quantum Optics},
Lecture Notes in Physics, (Springer, Berlin, 1993).
%
\bibitem{review} M.B. Plenio and P.L. Knight, Rev. Mod. Phys. {\bf 70}, 101 (1998).
\bibitem{Cohen-Tannoudji} C. Cohen-Tannoudji, J. Dupont-Roc, and G. Grynberg
{\em Atom-Photon Interactions} (John Wiley \& Sons Inc., 1992).
%
\bibitem{gant}
For the general case see e.g. F.R. Gantmacher, {\em Matrizentheorie}, Springer (Berlin 1986).
%
\bibitem{PLeniootro} G.C. Hegerfeldt and M.B. Plenio, Phys. Rev. A {\bf 53},
1164 (1996).
%
\bibitem{kim} The quantity $w_1(t,\psi_0)$ has been discussed in more detail for instance in
M.S. Kim, P.L. Knight, and K. Wodkiewicz, Opt. Comm. {\bf 62}, 385 (1987);
M.S. Kim and P.L. Knight, Phys. Rev. A {\bf 40}, 215 (1989).
%
\bibitem{quentin}
Q. Turchette et al., quant-ph/9806012. Submitted to Phys. Rev. Lett..
%
\end{references}
